\documentclass[11pt] {article}
\usepackage{changebar}
\usepackage{times}
\usepackage{graphicx}
\usepackage{makeidx}
\usepackage{natbib}
\usepackage{latexsym}
\usepackage{bm}
\usepackage{amsmath}
\usepackage{amsthm}
\usepackage{amssymb}
\makeindex

\newcommand{\beq}{\begin{equation}}
\newcommand{\eeq}{\end{equation}}
\newcommand{\dif}[2]{\frac{{\rm d} #1}{{\rm d} #2}}

\newcommand{\pdif}[2]{\frac{\partial #1}{\partial #2}}
\newcommand{\pddif}[3]{\frac{\partial^2 #1}{\partial #2 \partial #3}}

\newcommand{\defn}{\begin{quote}{\bf Definition. }}
\newcommand{\edefn}{\end{quote}}
\newcommand{\thm}{\begin{theorem}}
\newcommand{\ethm}{\end{theorem}}

\newcommand{\bmat}[1]{\left ( \begin{array}{#1}}
\newcommand{\emat}{\end{array}\right )}

\newcommand{\ts}{^{\sf T}} 

\newcommand{\bp}{{\bm \beta}}

\theoremstyle{definition}

\theoremstyle{plain}
\newtheorem{theorem}{Theorem}

\newcommand{\eps}[3]
{{\begin{center}
 \rotatebox{#1}{\scalebox{#2}{\includegraphics{#3}}}
 \end{center}}
}

\newcommand{\bfig}{\begin{figure}}

\begin{document}

\title{Inferring UK COVID-19 fatal infection trajectories from daily mortality data: were infections already in decline before the UK lockdowns? }

\author{ Simon N Wood {\tt simon.wood@ed.ac.uk} \\
School of Mathematics, University of Edinburgh, UK. }
\maketitle
\begin{abstract} The number of new infections per day is a key quantity for effective epidemic management. It can be estimated relatively directly by testing of random population samples. Without such direct epidemiological measurement, other approaches are required to infer whether the number of new cases is likely to be increasing or decreasing: for example, estimating the pathogen effective reproduction number, $R$, using data gathered from the clinical response to the disease. For Covid-19 (SARS-CoV-2) such $R$ estimation is heavily dependent on modelling assumptions, because the available clinical case data are opportunistic observational data subject to severe temporal confounding. Given this difficulty it is useful to retrospectively reconstruct the time course of infections from the least compromised available data, using minimal prior assumptions. A Bayesian inverse problem approach applied to UK data on first wave Covid-19 deaths and the disease duration distribution suggests that fatal infections were in decline before full UK lockdown (24 March 2020), and that fatal infections in Sweden started to decline only a day or two later. An analysis of UK data using the model of  \cite{flaxman2020lockdown}  gives the same result under relaxation of its prior assumptions on $R$, suggesting an enhanced role for non pharmaceutical interventions (NPIs) short of full lock down in the UK context. Similar patterns appear to have occurred in the subsequent two lockdowns. Estimates from the main UK Covid statistical surveillance surveys, available since original publication, support these results. Replication code for the paper is available in the supporting information of {\tt doi/10.1111/biom.13462}.
\end{abstract}

%



\section{Introduction \label{intro}}

Clinical data on the number of cases of Covid-19 (SARS-CoV-2) are subject to severe temporal confounding, as the rate of testing and criteria for testing have been changing rapidly on the same time scale as the infections, particularly in the early weeks and months of the epidemic. Because these are samples of convenience where the ascertainment fraction is changing and unknown, the data can clearly not be used to infer the actual number of infections. Neither, under normal circumstances, would statisticians recommend attempting to estimate the effective reproduction number of the pathogen from such data, since given the data problems the estimates must necessarily be driven strongly by the modelling assumptions \citep[see e.g.][\S 1.6 for general discussion of the problems with inference from non-random samples]{levine2001}. Indeed generically it is often very difficult to infer epidemiological parameters from clinical data, without the results being informed as much by the prior beliefs encoded in the model as by the data \citep[e.g.][]{lancet-ifr}. Much less problematic are estimates based on randomized surveillance testing, as now conducted in the UK by the office for national statistics (see Supporting Information for discussion of inferring incidence from testing data).

However some clinical data directly measure the quantity of epidemiological interest. This is the case for deaths with Covid-19 and for fatal disease duration. While not perfect, these data are less compromised than the data on cases. Deaths are reliably recorded and clinical grounds for suspecting Covid-19 are relatively clear for fatal cases, although accurately attributing death to a single cause is clearly not always possible. Good records are also often kept for such cases, with the result that there are several published studies on fatal disease duration \citep[][see section \ref{ddist}]{verity2020ifr,linton2020incubation,wu2020covid}. Although only possible with a delay of some weeks, it is of interest to establish what these relatively high quality data imply about the time course of infections, without strong modelling assumptions. 

Two types of daily death data are available. Daily reported deaths \citep[e.g.][]{worldometer} typically show marked weekly fluctuations as a result of weekly patterns in reporting delays, and may exclude deaths in some locations (such as nursing homes). Registered death data, such as the ONS data in the UK \citep{ons-deaths}, contain deaths in all locations and record exact date of death. \cite{nhs-covid-daily} publishes equivalent data for hospital deaths in England. The weekly cycle is less pronounced in these data, but their release is necessarily delayed relative to the daily reported deaths, although recent work partially overcomes this delay problem, by modelling the delays to enable `now-casting' of deaths by actual death date: see \cite{stoner2020deathcasting}. The right column of Figure \ref{infections} shows ONS, NHS and Swedish daily deaths by date of death (without now-casting).

The purpose of this paper is to show how a relatively straightforward statistical approach can be used to infer the fatal infection trajectory in the UK in a data driven way that makes the minimum of strong modelling assumptions. The approach is also applied to data from Sweden, the western European country offering the greatest policy contrast to the UK. Sweden never implemented full lockdown, sticking to less restrictive NPIs \citep[broadly aimed at `optimal mitigation' rather than `suppression' in the terms used by][who projected around 40 thousand deaths for this policy]{report12}. Meaningful quantification of the aggregate strength of restrictions that are intrinsically multivariate is difficult, but in terms of their aggregate economic impact, Swedish GDP dropped by about 2.9\% in 2020 compared to about 9.9\% for the UK.  Particular questions of interest are when the decline in fatal infections started in the UK and Sweden, whether UK infections were in substantial decline before full lockdown, whether the pathogen reproduction number was below 1 before lockdown, and how the timing of fatal incidence decline relates to the timing of the easing of lockdown.     

Answers to these questions may contribute to judging the proportionality of lockdown measures in the UK context, where there is strong statistical evidence for very large preventable life loss being associated with economic deprivation, and of economic deprivation being increased by economic shocks. This evidence is reviewed in detail in \cite{marmot-review-10}. For example the deprivation related life loss that the current UK population was due to suffer before the Covid crisis was 140-240 million life years \citep[or 2-3.5 years per capita, see][figure 2.3, for example]{marmot-review-10}. The range depends on whether the life expectancy of the lower decile or the lower half of the deprivation distribution is used as the reference for achievable life-expectancy. In examining the effects of the 2008 financial crisis and its aftermath, Marmot documents sharp reductions in life expectancy growth in the UK, which would imply a life loss burden in the 10s of millions of years. However attribution of such reduction-relative-to-trend is obviously very difficult. Less problematic is the 9 million life year loss implied by the increase in life expectancy gap between the more and less deprived halves of the UK population since 2008 \citep[7 weeks per capita, see][figure 2.5, for example]{marmot-review-10}: given the evidence presented in the review, this is more difficult to attribute to causes unrelated to the 2008 economic shock. The Bank of England estimates the shock to the UK economy caused by the response to Covid-19 to have been the largest for over 300 years, so there is a clear danger of substantial life loss being caused, given the historical data for the UK. For example, a feature of the 2008 crisis already repeated in 2020, is the reliance on a large programme of quantitative easing. Quantitative easing is credibly argued to directly increase economic inequality via mechanisms related to asset price inflation \citep[e.g.][]{fontan2016QE,domanski2016}.  There is some literature attributing some short-term life saving to recessions \citep[see e.g.][]{covid-life-loss-govt}, but the effects are modest relative to the long term effects reviewed by Marmot. 
For comparison with the above figures, the life loss that might have occurred from a minimally mitigated Covid-19 epidemic appears to be in the region of 3 million life years (2.5 weeks per capita). This is based on \cite{ons-lifetables} lifetables, ONS Covid-19 fatality by age data, a mid range infection fatality rate estimate of 0.006, a somewhat high herd immunity threshold of 0.7 and a 1 year lower bound life expectancy adjustment for co-morbidities based on \cite{hanlon2020}. It is broadly in line with the UK government estimates \citep{covid-life-loss-govt}. Given that 9 million life years, associated with the substantially smaller economic shock of 2008, is not negligible relative to 3 million life years potentially losable to Covid-19, there is obviously a delicate balance to be struck in the UK context, and evidence based on assumption light inference should probably play a role in shaping that balance. Another indicator of the difficulty of achieving the right balance is that the usual UK threshold for approving a pharmaceutical intervention is \pounds 30,000 per life year saved. On the basis of economic costs detailed in \cite{obrjuly20}, and the preceding life loss figures, the non-pharmaceutical interventions used in the UK appear to have a cost per life year saved that is an order of magnitude higher than this (excess government borrowing is projected to peak at \pounds 660 Billion in the OBR central scenario, for example). This discrepancy in willingness to pay  may lead to a problem of opportunity cost, as the same money can not be spent on preventing other life loss, such as that associated with economic hardship.

The remainder of the paper is structured as follows. Section \ref{ddist} discusses the available information on the distribution of fatal disease durations, and how to combine it while adequately characterizing the associated uncertainty. Section \ref{sec.models} introduces a simple generalized additive model for direct modelling of the daily deaths trajectories, and shows how it can be extended to infer the trajectory of fatal infections, either directly or by inferring the trajectory of the pathogen effective reproduction number, $R$, in a simple epidemic model. Since the extensions are not standard models, and are relatively expensive to compute with using standard Bayesian software, section \ref{sec.methods} outlines methods allowing computationally efficient inference with the models. Section \ref{sec:results} presents the main results on infection trajectories, and also the estimation of $R$. Section \ref{sec:check} discusses possible problems with the approach, in particular examining whether smoothness assumptions could be leading to substantial bias in inferred timings. Replication code and data are provided in the Supporting Information. 

\section{Fatal disease duration \label{ddist}}

Data on the incubation period from infection to onset of symptoms are analysed in many papers, for example \cite{lauer2020covid} found that the period is 2 to 11 days for 95\% of people, with a median of 5.2 days. A meta-analysis by \cite{McAloon20} suggests a log-normal distribution with log scale mean and standard deviation of 1.63 and 0.50. The uncertainty in this distribution is negligible in comparison to the uncertainty in the distribution of times from onset of symptoms to death discussed next. 

Several studies estimate the distribution of time from onset of symptoms to death, while properly controlling for the right truncation in the fatal duration data. \cite{verity2020ifr} found that the distribution of time from onset of symptoms to death for fatal cases can be modelled by a gamma density with mean 17.8 and standard deviation 8.44, based on 24 patients from Wuhan. \cite{wu2020covid} suggested a gamma density model with mean 20 and standard deviation 10 based on 41 patients from Wuhan. \cite{linton2020incubation} found that a log normal model offers a slightly better fit, and estimated a mean of 20.2 days and standard deviation of 11.6 days from 34 patients internationally. These distributions are shown in the left panel of Figure \ref{o2d}. A simple meta-analysis approach was used to combine the models. Samples of the correct size were simulated from each model and a log normal model was estimated by maximum likelihood for the combined resulting sample ($n=99$). A further log normal was also fitted (minimizing Kullback Leibler divergence) to the infection to death distribution implied by the fitted onset to death distribution and \cite{McAloon20} infection to  onset distribution (treated as independent). This process was repeated to generate replicate distributions. These replicate distributions were treated as draws from the distribution of infection to death distributions in subsequent analysis. 100 such draws are shown in fig \ref{o2d}.  The log normal was chosen because the careful analysis of \cite{linton2020incubation} found it to be a better model than the gamma. 

In addition, under strict conditions, I was able to access data on fatal disease durations for deaths occurring in English hospitals. Access to data with hospital acquired infections filtered out was not possible, so is was necessary to treat these data as a mixture of hospital and community acquired infections, as detailed in the Supporting Information. The resulting inferred fatal disease duration distribution for community acquired infection is plotted in blue in Figure \ref{o2d} and is consistent with the published studies.  

Finally, after the peer reviewed version of this paper was published, I realized that Figure 12 of  \cite{isaric.oct20} gives the cumulative distribution function of the time from hospitalization to death for 24421 patients in the ISARIC study who had died by the time of the report. Furthermore this is at a point in time at which recruitment to the study was at a low rate, so that right truncation problems should be relatively minor. The study also reports the mean and standard deviation of time from onset to hospitalization. Combining this information as described in the Supporting Information, gives the densities shown as dashed blue curves in  Figure \ref{o2d}. These data offer strong support for the distributions used in this paper, and in fact suggest that the results may be somewhat more certain than the plotted uncertainty bands imply.

\bfig

\eps{-90}{.67}{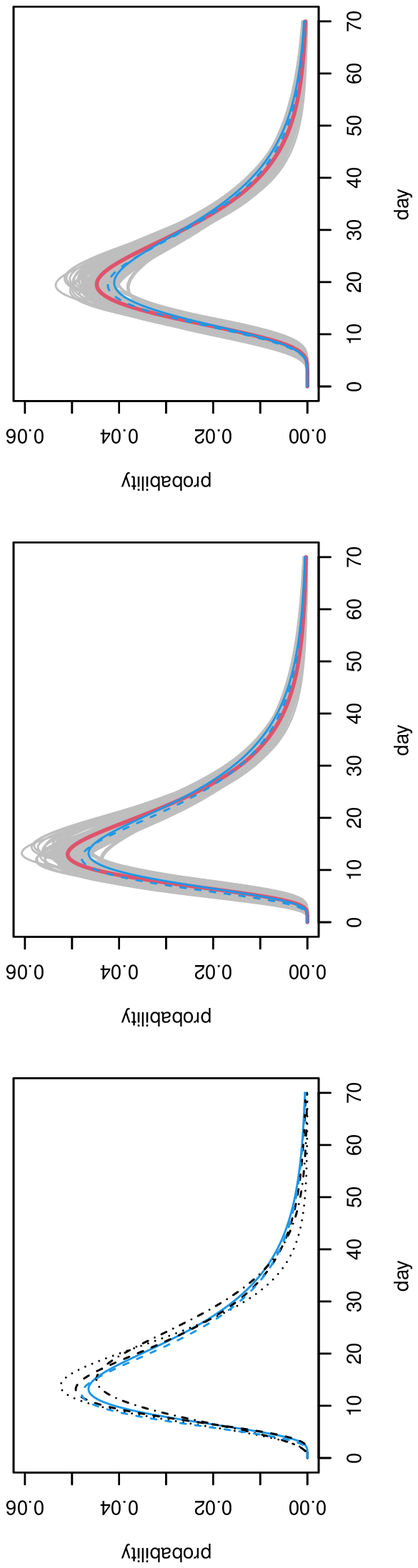}
\vspace*{-.5cm}
\caption{Fatal disease duration distributions. Left: onset to death. Dotted Verity {\em et al.} (2020) ; dashed Linton {\em et al.} (2020); dash-dot Wu {\em et al.} (2020); continuous blue is the log-normal mixture component for community acquired infection from the English hospital data; blue dashed is from the ISARIC study. Middle: combined Linton-Verity-Wu onset to death model, thick red is mean model, grey are 100 draws from the distribution of the combined model, thin blue are as left. Right: as middle, but combined infection to death model.}
\label{o2d}
\end{figure}

\section{Models \label{sec.models}}

This section first introduces a simple generalized additive model for daily death trajectories, and then shows how this can be extended to directly infer the trajectory of fatal infections (fatal incidence), without having to assume any particular dynamic model for the epidemic. The resulting model is no longer a generalized additive model and is the model that this paper advocates using. Its structure is such that any method for inference with the model can also be used for inference with the dynamic model of \cite{flaxman2020lockdown}, with appropriate restriction of the incidence trajectory to one representable with that model. The Flaxman model is presented to allow comparison of the results from the infection trajectory model with the apparently contradictory results of  Flaxman et al., but not to advocate its use.

{\bf Basic deaths series model}. Let $y_i$ denote the deaths or reported deaths on day $i$, assumed to follow a negative binomial distribution with mean $\mu_i$ and variance $\mu_i + \mu_i^2/\theta$. Let 
\beq
\log(\mu_i) = f(i) + f_w({\cal{D}}_i) \label{basic}
\eeq 
where $f $ is a smooth function of time measured in days, and $f_w$ is a zero mean cyclic smooth function of day of the week, ${\cal D}_i \in \{1,2,\ldots, 7\}$, set up so that $f^{[k]}_w(0) = f^{[k]}_w(7)$, where $k=0,1$ or 2  denotes order of derivative. $f(i)$ represents the underlying log death rate, while $f_w$ describes the weekly variation about that rate. The functions $f$ and $f_w$ can be represented using splines with associated smoothing penalties $\lambda \int f^{\prime\prime}(t)^2 dt$ and $\lambda_w \int f_w^{\prime\prime}({\cal D})^2 d{\cal D}$. Hyper-parameters $\lambda$ and $\lambda_w$ control the smoothness of the functions. The model is a straightforward generalized additive model and $(\lambda, \lambda_w)$ can be estimated as part of model fitting using a standard empirical Bayes approach as described in \cite{wood2017igam}. The model provides a good fit to both the reported deaths and ONS data. As expected $f_w$ is greatly attenuated for the ONS data (it vanishes for Swedish exact death date data). 

{\bf Infection trajectory model}. To estimate the daily infection trajectory the model is extended by expressing $f(i)$ in terms of the time course of earlier infections. Let $f_c(i)$ be the function describing the variation in the number of eventually fatal infections over time. Let $\bf B$ be the square matrix such that $B_{ij} = \pi(i-j+1;\mu,\sigma^2)$ if $i\ge j$ and 0 otherwise . $\pi$ denotes an infection-to-death log normal density as discussed above. For the moment its parameters, $\mu$ and $\sigma^2$, are treated as fixed but this will be relaxed in section \ref{sec:dist-unc}. Given the continuity of the log normal, the given form for $B_{ij}$ can be viewed as approximating an integral of $\pi$ over each day, using the midpoint of the integrand -- it is straightforward to approximate the integral more accurately, but given that $\pi$ is  originally estimated from durations discretized to whole days, any precision gain is illusory.  If ${\bf f}_c = [f_c(0),f_c(1), \ldots ]\ts$ and $ {\bm \delta} = [\delta(1), \delta(2),\ldots]\ts$ then ${\bm \delta} = {\bf Bf}_c$, where $\delta(i)$ is the expected number of deaths on day $i$. $\log f_c(i)$ can be represented using a spline basis, again with a cubic spline penalty. Working on the log scale ensures that $f_c$ is positive, but is also appealing because it means that a cubic spline penalty on $\log f_c(i)$ can be interpreted as a first derivative penalty $\int r^\prime(t)^2 dt$, acting on the epidemiologists `intrinsic growth rate', $r$. The final infection trajectory model is then obtained by simply substituting $f(i) = \log \delta(i)$ into (\ref{basic}). $\bf B$ is rank deficient, so inferring $f_c$ can be viewed as an inverse problem: without regularization multiple solutions that oscillate from day-to-day are possible. This ambiguity is removed by the smoothing penalty on $\log f_c$.  

{\bf Relaxed Flaxman model}. Since this work was originally undertaken in late April 2020, the work of \cite{flaxman2020lockdown} has appeared. Flaxman et al. make inferences about the reproduction number, $R$, and hence incidence rates, based on death trajectories and the fatal infection duration distribution of \cite{verity2020ifr}, but do so by modelling the pathogen effective reproduction number $R_t$ within a simple epidemic `renewal model'. \cite{flaxman2020lockdown} represent $R_t$ as a step function with steps allowed each time the government announced new containment interventions, and a sparsity prior promoting a small number of steps. In the notation of Flaxman {\em et al.} the expected number of infections each day (now total, rather than fatal) are denoted $c_t$. Given an initial $c_1$ the model is iterated from $t=2$ as follows
\beq
c_t = \left (1 - \sum_{i=1}^{t-1} c_i/N\right) R_t \sum_{\tau = 1}^{t-1} c_\tau g_{t-\tau} \label{epimod}
\eeq
where $N$ is the total initially susceptible population, $g_1 = \int_0^{1.5} \gamma(x) dx$ and $g_j = \int_{j-.5}^{j+.5} \gamma(x) dx$ for $j>1$. $\gamma$ is the p.d.f. of a Gamma distribution with shape parameter $6.5 \times 0.62^2$ and scale parameter $0.62^{-2}$. The $c_t$ values multiplied by the assumed infection fatality rate give ${\bf f}_c$. The level of the IFR only matters for the damping term in the first bracket of the expression for $c_t$ --- this has almost no effect in practice, a mid range value of 0.006 was used. The original assumptions about $R_t$ can be relaxed by representing $\log R_t$ using a spline basis, with associated penalty as for the other models, while $\log c_1$ is also treated as a free parameter. Hence $f_c$ in the infection trajectory model can simply be replaced by the Flaxman model with $\log R_t$ represented as a spline function. The model is otherwise unchanged. This model is presented only to allow comparison of this paper's results with those of \cite{flaxman2020lockdown}: its simple single compartment structure clearly does not meet the aim of inferring incidence with minimal assumptions. 

\section{Methods \label{sec.methods}}

The infection trajectory and Flaxman renewal models are not standard models estimable with standard software. They can  be implemented in Bayesian software, such as JAGS or STAN, but inference typically takes several hours if this is done. Dealing adequately with the uncertainty in the disease duration distribution multiplies this cost by 1-2 orders of magnitude. To avoid these problems an empirical Bayes approach can be employed.

\subsection{Basic inferential framework}

Direct inference about (\ref{basic}) uses the empirical Bayes approach of \cite{wood2015plig} in which the smooth functions are estimated by penalized likelihood maximisation \citep[e.g.][]{green.silverman}, with the smoothing parameters and $\theta$ estimated by Laplace approximate marginal likelihood maximization. Writing $\bp$ for the combined vector of  basis coefficients for $f$ and $f_w$, the penalized version of the log likelihood, $l(\bp),$ can be written
$$
l(\bp) - \frac{\lambda}{2} \int f^{[2]}(t)^2 dt - \frac{\lambda_w}{2} \int f^{[2]}_w({\cal D})^2 d{\cal D} = 
l(\bp) - \frac{1}{2} \bp \ts {\bf S}_\lambda \bp
$$
where ${\bf S}_\lambda = \lambda {\bf S}_f + \lambda_w {\bf S}_w$: ${\bf S}_{f}$ and ${\bf S}_{w}$ are known constant positive semi-definite matrices. Smoothing parameters, $\lambda$ and $\lambda_w$, control the smoothness of $f $ and $f_w$. Let $\widehat \bp$ be the maximizer of the penalized log likelihood, and $\bf H$ its negative Hessian at $\widehat \bp$. Viewing the penalty as being induced by an improper Gaussian prior, $\bp \sim N({\bf 0}, {\bf S}_\lambda^-)$, $\widehat \bp$ is also the MAP estimate of $\bp$. Furthermore in the large sample limit 
\beq
\bp |{\bf y} \sim N(\widehat \bp,({\bf H} + {\bf S}_\lambda)^{-1}). \label{gg}
\eeq 
Writing the density in (\ref{gg}) as $\pi_g$, and the joint density of ${\bf y} $ and $\bp $ as $\pi({\bf y}, \bp)$, the Laplace approximation to the marginal likelihood for the smoothing parameters $\bm \lambda$ and $\theta$ is $\pi({\bm \lambda},\theta) =\pi({\bf y}, \bp)/\pi_g(\bp|{\bf y})$. Nested Newton iterations are used to find the values of $\log({\bm \lambda}),\theta$ maximizing $\pi({\bm \lambda},\theta)$ and the corresponding $\widehat \bp$ \citep[for details see][]{wood2015plig}.
 
Given (\ref{gg}) credible intervals for $f$ are readily computed, but it is also straightforward to make inferences about when the peak in $f$ occurs. Simply simulate replicate coefficient vectors from (\ref{gg}) and find the day of occurrence of the peak for each corresponding underlying death rate function, $f$. 

\subsection{Extension for the infection and Flaxman models}

While inference about (\ref{basic}) using the preceding framework requires little more than a call to the {\tt gam} function in R package {\tt mgcv}, its application to the other models, which are not generalized additive models, requires more work. 
For the  model formulated in terms of $f_c$ this requires expressions for the negative binomial deviance (or log likelihood) and its derivative vector and Hessian matrix w.r.t. the model coefficients. 

First consider the negative binomial deviance for observation $i$,
$$
D_i = 2y_i\log\{\max(1,y_i)/\mu_i\} - (y_i + \theta) \log\{(y_i + \theta)/(\mu_i +\theta) \},
$$ 
$$
\dif{D_i}{\mu_i} = 2\left (\frac{y_i + \theta}{\mu_i + \theta} - \frac{y_i}{\mu_i} \right)
~~~\text{and}~~~ \dif{^2 D_i}{\mu_i^2} = 2 \left ( 
\frac{y_i}{\mu_i^2} - \frac{y_i+\theta}{(\mu_i + \theta)^2}
\right ).
$$
These need to be transformed into derivatives w.r.t. $\bm \beta$, as follows:
$$
\pdif{D_i}{\beta_j} = \dif{D_i}{\mu_i}\pdif{\mu_i}{\beta_j} ~\text{and}~
\pddif{D_i}{\beta_j}{\beta_k} = \dif{^2 D_i}{\mu_i^2} \pdif{\mu_i}{\beta_j}\pdif{\mu_i}{\beta_k} +
\dif{D_i}{\mu_i} \pddif{\mu_i}{\beta_j}{\beta_k}.
$$
Writing ${\bf X}^f$ and ${\bf X}^w$ for the model matrices for the smooth terms $\log f_c$ and $f_w$, we have ${\bm \delta} =  {\bf Bf}_c $ where ${\bf f}_c = \exp({\bf X}^f{\bm \beta}^f)$ (here $\exp(\cdot)$, division and multiplication are applied element-wise to vectors), and ${\bf f}_w = {\bf X}^w{\bm \beta}^w$. Then ${\bm \mu} = \exp(\log {\bm \delta} + {\bf f}_w )$, while 
$$
\pdif{\bm \mu^{~}}{{\bm \beta}^f} = \text{diag}({\bm \mu}/{\bm \delta}){\bf B} \pdif{{\bf f}_c}{{\bm \beta}^f}, ~~~
\pdif{\bm \mu^{~}}{{\bm \beta}^w} = \text{diag}({\bm \mu}) {\bf X}^w,
$$
$$
\pddif{{\bm \mu}}{\beta_j^w}{\beta_k^w} = {\bm \mu} {\bf X}^w_{\cdot,j} {\bf X}^w_{\cdot,k}, ~~~
\pddif{{\bm \mu}}{\beta_j^f}{\beta_k^f} = \text{diag}({\bm \mu}/{\bm \delta}){\bf B}
\pddif{{\bf f}_c}{\beta_j^f}{\beta_k^f} 
$$
$$
\text{~and~}
\pddif{{\bm \mu}}{\beta_j^f}{\beta_k^w} = \text{diag}({\bf X}^w_{\cdot,k}{\bm \mu}/{\bm \delta})
{\bf B} \pdif{{\bf f}_c}{{\bm \beta}^f} .
$$
For the given representation of ${\bf f}_c$
$$
\pdif{{\bf f}_c}{{\bm \beta}^f} = \text{diag}({\bf f}_c) {\bf X}^f ~~ \text{and} ~~
\pddif{{\bf f}_c}{\beta_j^f}{\beta_k^f} = \text{diag}({\bf f }_c) {\bf X}^f_{\cdot,j} {\bf X}^f_{\cdot,k}.
$$

When using the relaxed Flaxman model, the preceding derivatives of ${\bf f}_c$ have to be replaced with derivatives of ${\bf f}_c$ w.r.t. the coefficients of the spline representing $\log R_t$. Routine application of the chain rule to (\ref{epimod}) gives corresponding iterations for the derivatives of $c_t$, and hence ${\bf f}_c$, w.r.t these spline coefficients and $\log c_1$.  

Given these expressions and the penalties, $\widehat \bp $ can be obtained by Newton iteration, given smoothing parameters. To estimate smoothing parameters, the simplest approach is to fix the negative binomial $\theta$ at its estimate from model (\ref{basic}), and use \cite{wood2016gfs}, alternating generalized Fellner Schall updates of the smoothing parameters  with updates of $\widehat \bp $ given those smoothing parameters.  This finds the smoothing parameters to approximately maximise the model marginal likelihood. The non-linearity of the renewal equation model means that some effort is required to get non-absurd starting values. I got these by a few minutes of experimentation with simple step functions for the initial $\log R_t$ to get death trajectories of roughly the shape and amplitude of the true trajectories (a close initial fit is not required: initial deviances 2 orders of magnitude greater than for the final fit were unproblematic).   

Given $\theta$ and the smoothing parameters, the approximate posterior (\ref{gg}) could be used directly, or as the basis for the proposal distribution in a simple Metropolis Hastings sampler. A fairly efficient sampler results from alternating fixed proposals based on (\ref{gg}) with random walk proposals based on zero mean Gaussian steps with a shrunken version of the posterior covariance matrix. By this method, effective sample sizes $>5000$ for each coefficient took about 40 seconds computing on a low specification laptop. This was the approach used for the infection trajectory model. The results were indistinguishable from those produced at the cost of several hours of computing using JAGS \citep{plummer2003jags,coda} to simulate from the model posterior. 

\subsection{Disease duration distribution uncertainty \label{sec:dist-unc}}

The methods so far perform inference conditional on fixed values for the parameters $\mu$ and $\sigma^2$ of the log normal density describing the infection to death duration distribution. In reality there is uncertainty about these parameters. To incorporate this uncertainty into the infection trajectory model, inference was run for each of the 100 sample distributions shown in grey in the right hand panel of Figure \ref{o2d}, and the resulting posterior samples were pooled, to give a sample from the unconditional posterior distribution of the model.

\section{Results \label{sec:results}}

\bfig

\eps{-90}{.58}{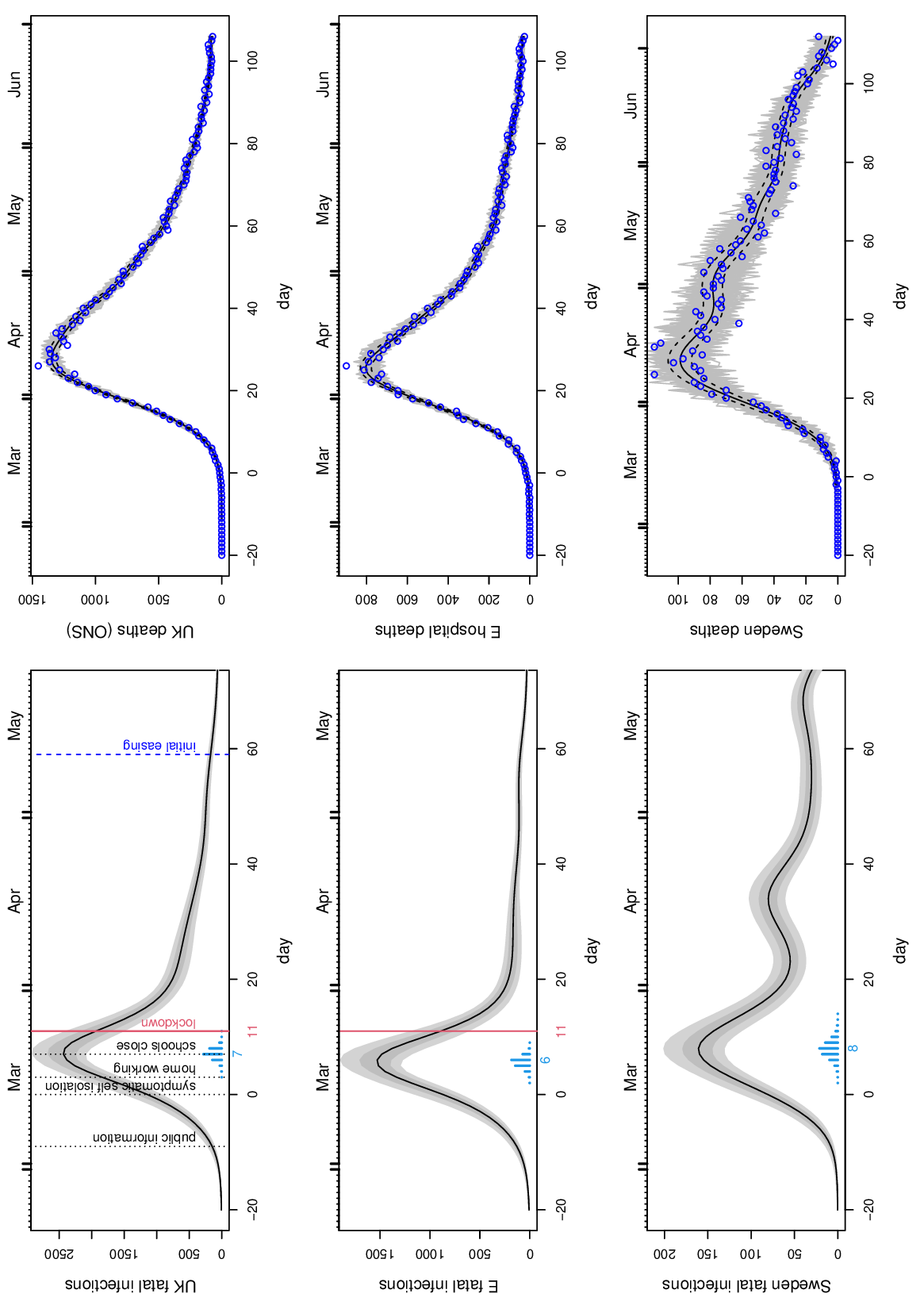}
\caption{ In all plots black curves show  the posterior median while light grey and dark grey regions show respectively 95\% and 68\% confidence regions, including uncertainty in the fatal disease duration distribution. Day 0 is 13th March 2020, and the vertical red line marks the first day of UK lockdown. Top left: Inferred daily fatal infection rate, $f_c$, for the UK.  The scaled barchart shows the posterior distribution for day of peak infection with the peak day labelled. NPI start dates are marked by labelled vertical lines. Top right: Consistency check. In grey are 100 sets of death data simulated forward from the inferred median fatal infection profile. Symbols are the ONS daily death data for the UK on which inference is based. The dashed curves are 95\% confidence intervals for underlying death rate estimated by direct fitting of (\ref{basic}). 
Middle row: As top row, but using the NHS England daily hospital death data. Note that the inferred infection trajectories are substantially different to time lagged versions of the deaths trajectories.  
Bottom row: as the previous rows, but for Sweden.
 \label{infections}}
\end{figure}


Figure \ref{infections} shows the results of applying the model to the Office for National Statistics daily Covid-19 death data for the UK, to the NHS England hospital data and to the daily death data for Sweden from \cite{sweden-covid-daily}. The results include the uncertainty about the disease duration distribution shape.   ONS and NHS data are up to 27th June -- including later data simply narrows the uncertainty, while making negligible difference to the overall conclusions. The most notable feature of the results is that fatal infections are inferred to be in substantial decline before full lockdown (the same result was obtained by this method in early May 2020, based on the first 50 days of reported daily death data). Sweden appears most likely to have peaked only one or two days later (barring some systematic difference in fatal disease durations for Sweden), having introduced NPIs well short of full lockdown. The results also emphasise the fact that the infection trajectory is not simply a time shifted version of the death trajectory (assuming it was might lead to unwarranted delay in easing lockdown, for example). The difference in timing and shape of the inferred profile between the ONS and NHS data reflects the fact that the latter contain care home data. There is an argument for preferring hospital data for inferring community fatal infections, in that the care home epidemic is now known to have special features with at least some of the infection not coming from normal community transmission. See in particular \cite{comas2020care-home} for a discussion of care home deaths internationally, including the UK. In addition, in the UK, care home deaths were often attributed to Covid-19 without a test, especially after death certification guidelines were changed to encourage reporting of suspected, rather than confirmed Covid-19 deaths. The care home data therefore have some under-reporting of Covid deaths, followed by over-reporting (the signal of this is visible in ONS data in the change in non-Covid pneumonia deaths being reported, relative to normal, for example).

\bfig

\eps{-90}{.55}{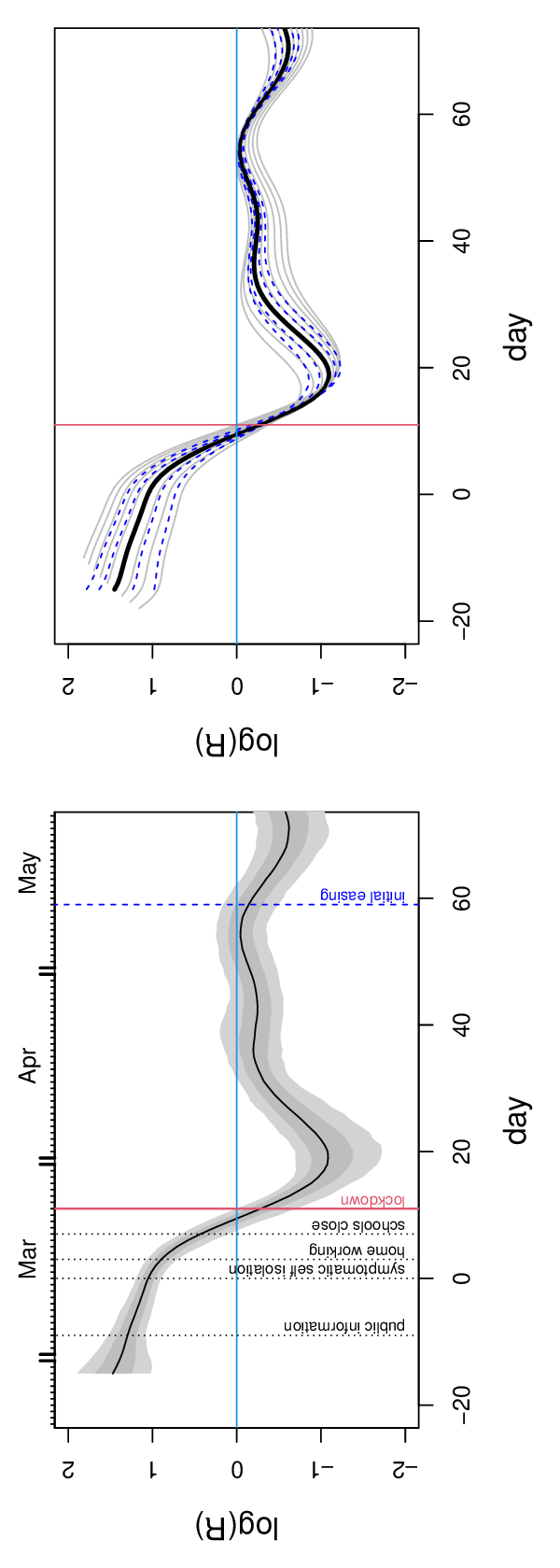}

\caption{ Left: Estimates and confidence bands for the effective reproduction number, $R$, from a simple SEIR model given the inferred infection profile (incidence), $f_c$.  The assumed mean time to infectivity was $1/\gamma = 3$ days and the mean infectivity duration was $1/\delta = 5$ days. The labelled vertical bars show policy change dates in March 2020. 
Given the rapidity of policy change relative to the epidemic's dynamic time scale, and government policy sometimes lagging behaviour, casual over interpretation of these timings should be avoided. Right: sensitivity analysis. Dashed blue -- time to infectivity was varied from 1 to 5 days. Grey -- duration of infectivity was varied from 2 to 10 days. Logs are natural. $R$ appears to be below 1 before full lockdown, but fell further after it. 
 \label{Rfig}}
\end{figure}

Taken together the results for the UK and Sweden raise the questions of firstly whether full lockdown was necessary to bring infections under control, or whether more limited measures might have been effective, and secondly whether the several month duration of full lockdown was appropriate. These emphasise the desirability of statistically well founded direct measurement of epidemic size through randomized testing. Had such testing being carried out leading up to lockdown it would have been clearer if the measures preceding lockdown (see Figures \ref{infections} and \ref{Rfig}
)  were working, or whether stronger restrictions were needed. Similarly such testing might have given earlier indication of when lockdown could be eased. Instead management was reliant on a complex modelling synthesis of expert judgement and problematic clinical case data. Less statistically problematic reconstructions, like the one presented here, are clearly only possible weeks after the fact. Note that while it is natural to interpret these fatal infection trajectories as proportional to the overall infection trajectories, that will only be the case if the infection fatality rate is constant over time. There is evidence for improvements in hospital care from late March onwards that suggest that this is might not be the case \citep[see][]{dennis2021covid-improve}. The Supporting Information includes a sensitivity analysis of this issue: it has the potential to right shift the peak incidence by up to a day and to lead to somewhat less rapid decay of the incidence trajectory. 


\subsection{Inferring R}

Much public debate has focused on the effective reproduction number, $R$, and in theory it is possible for a decline in the rate of infections to be only temporary as a result of $R$ dropping but remaining above one. Could it be that the declines in $f_c$ seen before lockdown were of this short term type, and that renewed increase would therefore have occurred without full lockdown? The answer appears to be no. $R$ is all but impossible to measure directly, so inference about it requires assumption of an epidemic model. However, given an epidemic model it can be directly inferred from the reconstructed infection profile. For example consider a simple SEIR model: $\dot{S} = -\beta S I$, $\dot{E} = \beta S I - \gamma E$, $\dot{I} = \gamma E - \delta I$ (here $\delta I $ is the rate of recovery {\em or} progression to serious disease). $\widehat f_c$ is a direct estimate of $\beta S I$ (to within a constant of proportionality), so by solving  $$\dot{E} = \widehat f_c - \gamma E, ~~~~~~\dot{I} = \gamma E - \delta I $$ (from 0 initial conditions) the direct estimate $R = f_c/(I\delta)$ is readily computed (any constant of proportionality cancels in $R$). A different epidemic model could be used here of course: see \cite{diekmann1990R} for calculation of $R$ in general from a model. Figure \ref{Rfig} shows the results using $\widehat f_c$ for the English hospital data for plausible values of average time to infectivity of $1/\gamma = 3$ days and mean duration of infectiousness of $1/\delta = 5$ days, along with sensitivity analysis for these values. The credible intervals shown include the uncertainty about the fatal disease duration distribution. $R $ appears to be below 1 before full lockdown. 

A useful feature of the $R$ estimates is to emphasise that the analysis in this paper in no way suggests that lockdown did not have an effect on transmission. Even if $R$ was below one before lockdown, full lockdown can only have reduced it further, and the estimates in Figure \ref{Rfig} are obviously consistent with this. Note, however that the recovery in $R$ after the post lockdown dip is to be expected, given the simple fact that $R$ is the number of new infections created per infection, {\em averaged over the population of infections}, not the population of people. Broadly speaking, at lockdown the population of people, and infections, was split into the locked down population, where infections could create few new infections, and the `unlocked' population where the reproductive rate of the pathogen was higher (assuming lockdown had an effect). An initial average over all infections is then dominated by those infections in the locked down population, giving a low $R$ (especially once the possibilities for infecting locked down household members have been exhausted). As the infections in the locked down population die out, the proportion of all infections that are in the unlocked population must increase -- so that an average over all infections must yield a higher $R$ again.

\subsection{The Flaxman model}

\bfig

\eps{-90}{.55}{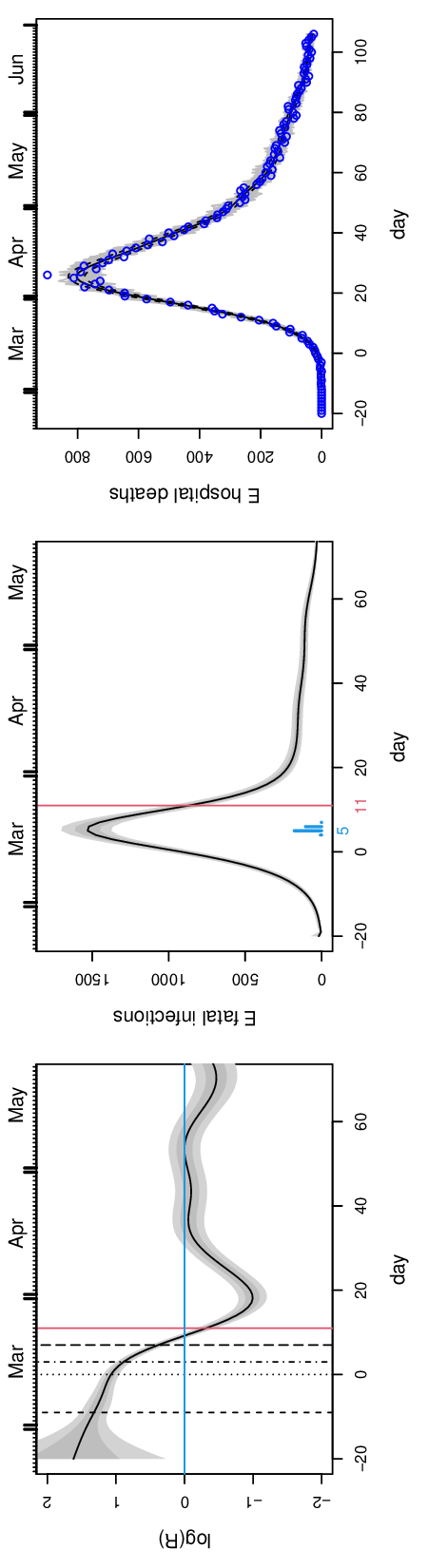}

\caption{Results from the epidemic model of Flaxman {\em et al.} (2020), with the assumptions on $R$ relaxed: $\log R$ is assumed smooth and continuous. Left: the inferred $R$ from fitting the NHS hospital data. The inferred $R$ trajectory is similar to the one shown in Figure \ref{Rfig}, despite the different model structure. Intervals do not include disease duration distribution uncertainty here. Middle: the corresponding fatal infection profile. Right: the simple sanity check as in Figure \ref{infections}.
 \label{renewal}}
\end{figure}

As noted above in section \ref{sec.models}, \cite{flaxman2020lockdown} also analysed death trajectories, using a simple epidemic model, but came to conclusions apparently contradicting Figure \ref{Rfig}. They  concluded that only after full lockdown did $R$ drop below 1, and that fatal infections continued to increase up until the eve of full lockdown. \cite{flaxman2020lockdown} used the \cite{verity2020ifr} fatal disease duration distribution, so the difference in results does not lie there. To describe the epidemic dynamics Flaxman {\em et al.} use the simple single compartment discrete renewal model (\ref{epimod}). Within that model they assume that $R$ is constant between the imposition of interventions, but can undergo a step change at each intervention: the steps are free model parameters. This model for $R$ is quite restrictive.  In particular it does not  allow $R$ to change after lockdown, despite the fact that at lockdown the population has been stratified in a way that the renewal model does not represent, so that some compensating flexibility in $R$ is likely to be required to avoid modelling artefacts. At the same time the model is rather underdetermined preceding lockdown, because of the frequent intervention changes. This indeterminacy in the model is addressed by using a sparsity promoting prior on the step changes in $R$, which favours few larger changes, rather than several smaller changes (see the supplementary material for Flaxman {\em et al.} for a description of this prior). When using the model to simultaneously model multiple European countries there is a further assumption that the intervention effects are the same for all countries (despite the different order of their implementation) and that only the lockdown effect varies between countries. It seems likely to be difficult to pick up effects of the interventions preceding lockdown from such a model structure.   

A relaxed version of the Flaxman model in which $\log R_t$ is a continuous function is described in section \ref{sec.models}. The results from using this model for inference using the NHS hospital data are shown in Figure \ref{renewal}. The relaxation of the assumptions on $R$ brings the results (for the UK) into alignment with those in the rest of this paper, and into broad  consistency with developments later in the year, which are otherwise difficult to square with \cite{flaxman2020lockdown}.


\subsection{Later infection waves}

While the initial motivation for this work was to provide reasonably timely analysis for the first wave, based on the limited data available by May 2020, the methods scale readily to the much longer data runs available by early 2021. The only change is that it makes sense to use an adaptive smoother \citep[see e.g.][\S 5.3.5]{wood2017igam} for $f(t)$, in which the degree of smoothness is allowed to vary with time. The longer data runs make it feasible to estimate the multiple smoothing parameters that this entails. Using an adaptive smooth guards against artefacts driven by the smoothness that is appropriate on average, for all the data, not being appropriate at times of rapid change.

 \bfig

\eps{-90}{.55}{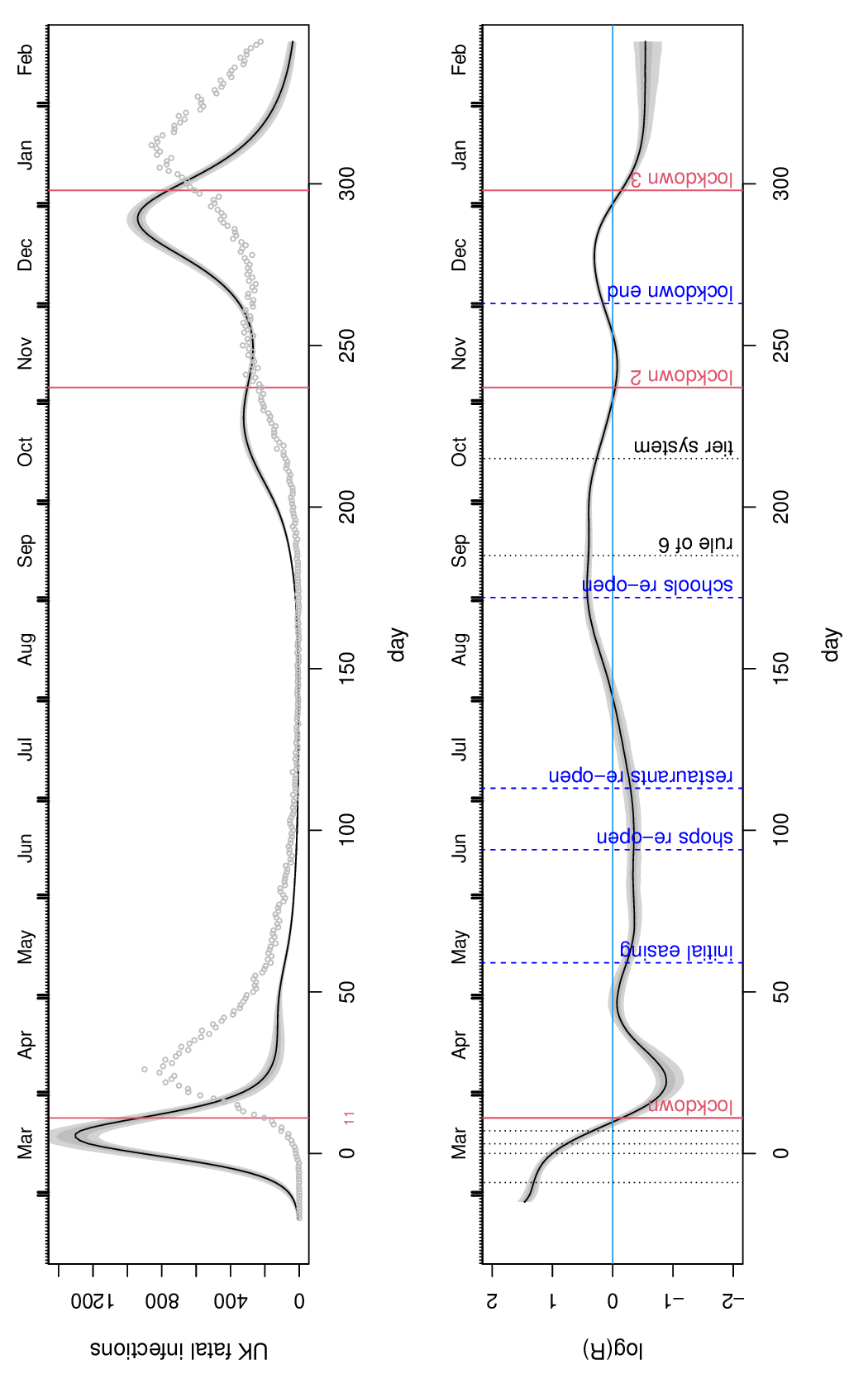}

\caption{Inference for the English hospital deaths data up to mid February 2021, including disease duration uncertainty. Top: Inferred fatal incidence. Grey symbols are the hospital deaths from which incidence is inferred. Red vertical lines mark the start of each of the three English lockdowns. Note that improvements in medical treatment mean that the IFR is very likely not to be constant between the first and later waves, so comparing their relative sizes is difficult. Bottom: the inferred $R$ using the simple SEIR approach. NPI impositions short of lockdown are marked by dotted vertical lines, relaxations are marked by dashed lines.
 \label{epidemic}}
\end{figure}

The results of this application are shown in Figure \ref{epidemic}. Note that likely changes in infection fatality rate as a result of improved hospital treatment mean that the relative sizes of the fatal infection incidence curves in the first and subsequent waves can not be interpreted as reflecting the relative sizes of total incidence (the later incidence curves would need to be scaled up somewhat). Causal over-interpretation 
of the $R$ curves should be avoided, not least because there is no reason to expect Covid-19 not to display the seasonality in transmission common to other respiratory illnesses. However, the results are obviously inconsistent with full lockdowns having caused $R<1$, since cause should not happen after effect. Further, the drop in $R$ seen after the initial NPIs were introduced, but before full lockdown, does seem consistent with the levels of $R$ later achieved while measures short of lockdown were in place. The interesting feature of $R$ apparently increasing from quite early in the second lockdown, might relate to the spread of the new variant, but of course also occurs at a time when respiratory infections generally start to increase. Likewise the further increase until mid December, may well be due to the new variant, but increased activity in the run up to Christmas is also likely to be a factor -- incidence appears to peak over the Christmas to New Year period. Vaccine rollout seems virtually certain to be a major factor in pushing down $R$ and fatal incidence from December. The vaccine has been given to those most at risk first, so the constant IFR assumption required to interpret fatal incidence as proportional to total incidence obviously no longer holds. This further implies  that the inferred $R$ is in some sense only the $R$ relevant to the `at serious risk' population. Of course, it could be argued that for epidemic management purposes, the fatal incidence and the corresponding $R$ are of primary interest. 

Interestingly the pattern observed at the second lockdown and in the preceding months is consistent with the results reported by \cite{rep41} who analysed regionally stratified death, hospital occupancy and testing data for 2020 up until December, using a highly detailed age structured SEIR with added health service compartments. The entire trajectory up until December is also consistent with the results of \cite{wood2021rep41}, who re-implemented the Knock et al. model, but removed some of its very strong modelling assumptions around the first lockdown.

\section{Model checking \label{sec:check}}    

While standard residual checks indicate no problem with the model from the point of view of statistical fit, there are three issues which could potentially undermine the results, and a further issue relating to interpretation. 

The first relates to the infection to death interval distribution and the fact that the death data contain an unknown proportion of patients whose infection was hospital acquired. These patients are likely to have had shorter disease durations, since they were already sufficiently unwell or frail to be in hospital. This paper has inferred when the fatal infections would have occurred if they were all community generated, since it is the community infections that are of interest with respect to the effects of lockdown, social distancing etc. Without knowing even the proportion of deaths from hospital acquired infection it is anyway not possible to do otherwise. 

The presence of hospital infections in the death data will bias inference about the dynamics of community fatal infections if it substantially changes the shape of the deaths profile, relative to what would have occurred without hospital infection. Broadly, if the trajectory of hospital acquired infection deaths peaked earlier than the overall trajectory, then the community infection peak will be estimated to be earlier than it should be (since the true community infection death peak is then later). Conversely, if the hospital acquired infection deaths peaked later, then the community infection peak will be estimated as being later than it should be. The degree of bias will depend on the proportion of hospital acquired infections and the degree of mismatch in timings. It is difficult to judge which alternative is more likely: standard epidemiological modelling assumptions would imply that the more community acquired cases are hospitalised the more hospital infections would occur and that hospital infections will lag community cases. But against this, hospital acquired fatal disease durations are likely to contain a higher proportion of shorter durations. In any case the proportion of hospital acquired infections in the death series would have to be quite high for the issue to substantially modify the conclusions. 

The second issue is that age dependency in the duration distribution coupled with shifts in the age structure of deaths over time could also be problematic. However, as documented in the Supporting Information, the data for England and Wales show remarkably little variation in the age structure of Covid-19 fatalities over the course of 2020, while analysis of English hospital data apparently shows little evidence for age dependence in the disease duration distribution.  


\bfig

\vspace*{-.6cm }

\eps{-90}{.55}{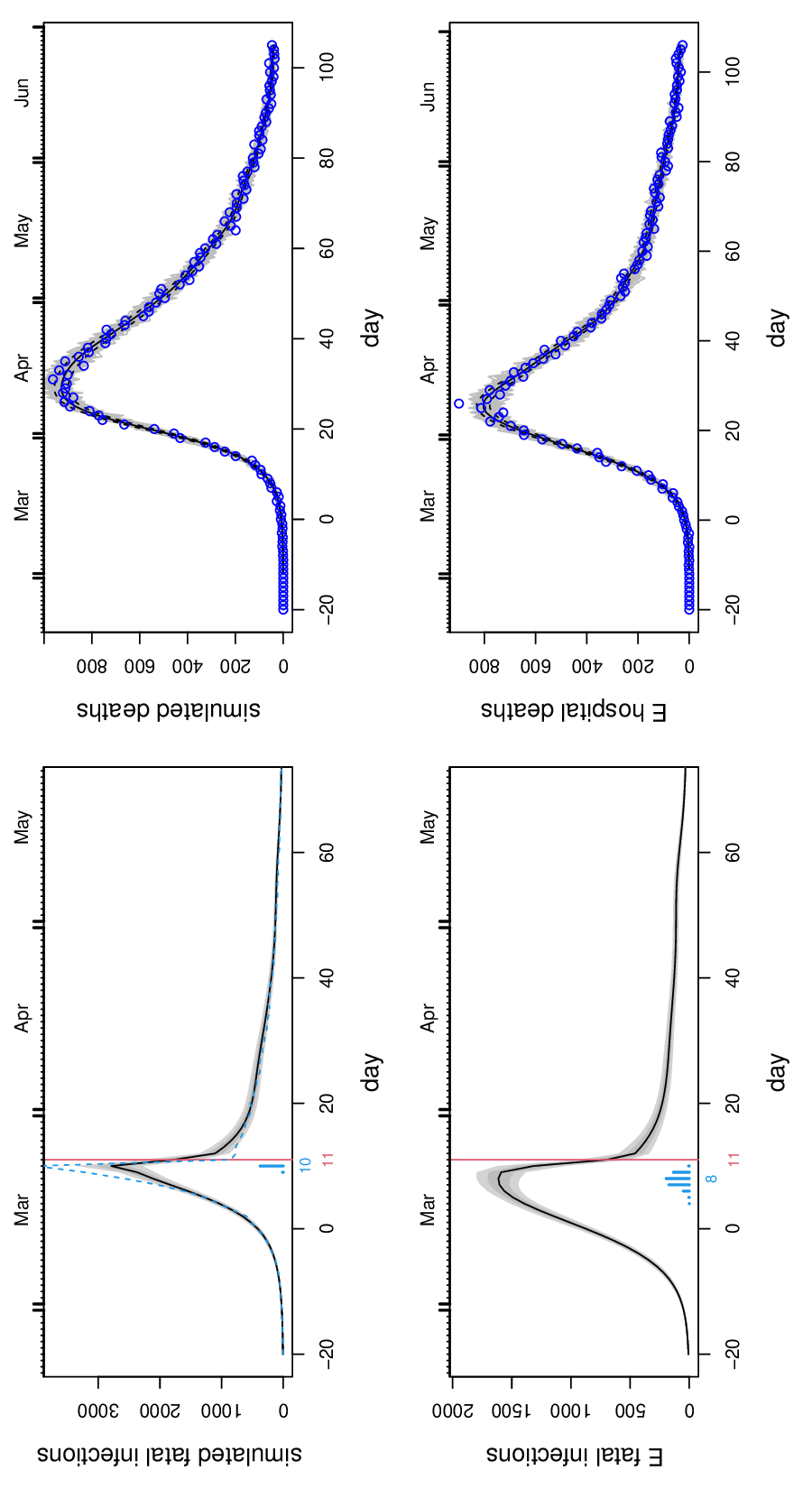}

\vspace*{-.5cm}

\caption{Model checking plots in which the smoothness assumptions are relaxed around lockdown by a time dilation, in order to allow accurate capture of any extremely discontinuous infection profile in this region. The top row shows the method reconstructing an extreme simulation scenario in which there was no reduction in transmission rate up until lockdown, and then an instantaneous drop. Left: the reconstruction (plot meaning as Figure \ref{infections}) with the true simulated daily infections shown dashed. Right: forward simulation from the median profile as in Figure \ref{infections}. The blue symbols are the simulated death data used for inference. The bottom row is for the NHS England hospital data under the time dilated model. Even this model deliberately modified to promote a very abrupt change at lockdown suggests that the infection rate was probably declining before lockdown.
\label{dilation}}
\end{figure}

The third issue is whether the smoothing penalty on $\log f_c$ would lead to systematic mis-timing of the estimated peak under the scenario of a very asymmetric peak in the true infection profile around lockdown. To investigate this, data were simulated from a model in which the underlying infection rate increased geometrically, doubling every 3 days until lockdown, when the rate dropped immediately to 0.2 of its peak value, shrinking thereafter by 5\% per day. Fatal infections were simulated as Poisson deviates with the given underlying rate. This model is an extreme scenario, in which measures prior to full lockdown had no effect, and the effect of lockdown was instant, as if the locked down population (i.e. those not in essential work) had isolated alone, rather than increasing their contact with members of their household while drastically reducing it with everyone else. However it is the scenario implicit in much public discussion in the UK, at least at the time that this work was originally conducted. Under this scenario, the method does indeed tend to incorrectly estimate the infection peak as 2 to 3 days before lockdown, rather than the day before, as it struggles to accommodate the drop.

The naive approach to this issue is to introduce a parameter at lockdown representing an instantaneous drop in infections. However doing so introduces a very strong structural assumption into the model, undermining the aim of avoiding strong assumptions. This approach also has the serious side effect of introducing non-parametric smoothing boundary effects on both sides of the break. These boundary effects severely compromise inference in the most interesting region of the infection profile, while simultaneously increasing the importance of the structural assumption at the expense of the data. Indeed when such a model is built it estimates a large drop even from data simulated from a smooth infection profile. It also estimates such a drop if the drop's location is moved (for simulated or real data). 

A better approach is to use a smooth time-dilation to relax, but not eliminate, the model smoothness assumptions in the vicinity of lockdown. The dilation is made sufficient that the model can accurately  capture the extreme scenario in the simulation, but without imposing a break and boundary effects. In particular $f_c$ and its smoothing penalty are computed with respect to a version of time which makes the day before, of and after lockdown count as 3.5, 6 and 3.5 days, respectively. Obviously regular un-dilated time is used for mapping infections to deaths. For the extreme simulation, the model then correctly gives most posterior probability to the day before lockdown as the peak. In contrast the same model for the real data has very low probability of the peak being the day before lockdown rather than earlier. 

Figure \ref{dilation} shows the results from fitting the time dilated model to the extreme simulation scenario and to the NHS England hospital data. Even this model, deliberately modified to favour a very abrupt change at lockdown, suggests that infections started to decline before lockdown, with the most likely day for the peak only 1 day later than with the un-dilated model. The Supporting Information includes similar checks for the \cite{flaxman2020lockdown} model, with similar conclusions.
 
 Finally, interpretation of the fatal incidence trajectories as proportional to the overall incidence trajectories rests on an assumption that the infection fatality rate is constant over time. There is evidence that the hospitalized case fatality rate declined in the two months or so after the peak of the first wave of infections \citep{dennis2021covid-improve}, with this effect not explicable by any detectable change in patient characteristics. However, on the ground changes in the severity threshold for admission would be very difficult to detect, seem likely at times when some hospital's were at or near capacity, and could also contribute to such a pattern. The Supporting Information includes a check of the impact that the reported improvements would have on the shape of inferred overall incidence. The peak incidence could be shifted by as much as a day later, and there would be a somewhat slower decline in incidence relative to the results plotted in Figure \ref{infections}.

\section{Discussion \label{disc}}    

The analysis in this paper does not absolutely prove that substantial decline in fatal infections in the UK preceded the first full lockdown by several days, but it very strongly suggests that this is what happened, and the ISARIC fatal disease duration distribution data, not available in the peer reviewed version of the paper, only strengthens this point further. Since the peer reviewed publication, information has also become available from the UK's two main Covid surveys based on proper randomized statistical sampling. The Office for National Statistics PCR prevalence survey has now published its estimates of incidence. For the second and third lockdown these support the analysis in this paper. See Figure \ref{ons}. Furthermore the REACT-2 survey \citep{ward2021react} has reconstructed numbers of newly symptomatic cases per day from survey participants testing positive for Covid-19 antibodies. Lagging these by the published mean time from infection to first symptoms gives the left plot of figure \ref{ons}. This agrees with the results of this paper for all three lockdowns. Some differences in relative peak sizes between the series are to be expected if there are changes in infection fatality rate, and as a result of the vaccination program. Taken together these results, from the 3 highest quality data sources available for the UK epidemic, leave little room for reasonable doubt that incidence was in decline before each lockdown. Apparent demonstrations of the contrary result appear to rely on building it into the modelling assumptions, or on highly informal reasoning about timings or correlations, or on a view that scientific opinion, in sufficient quantity, outweighs data and measurement. 

 \bfig

\vspace*{-.6cm }

\eps{-90}{.55}{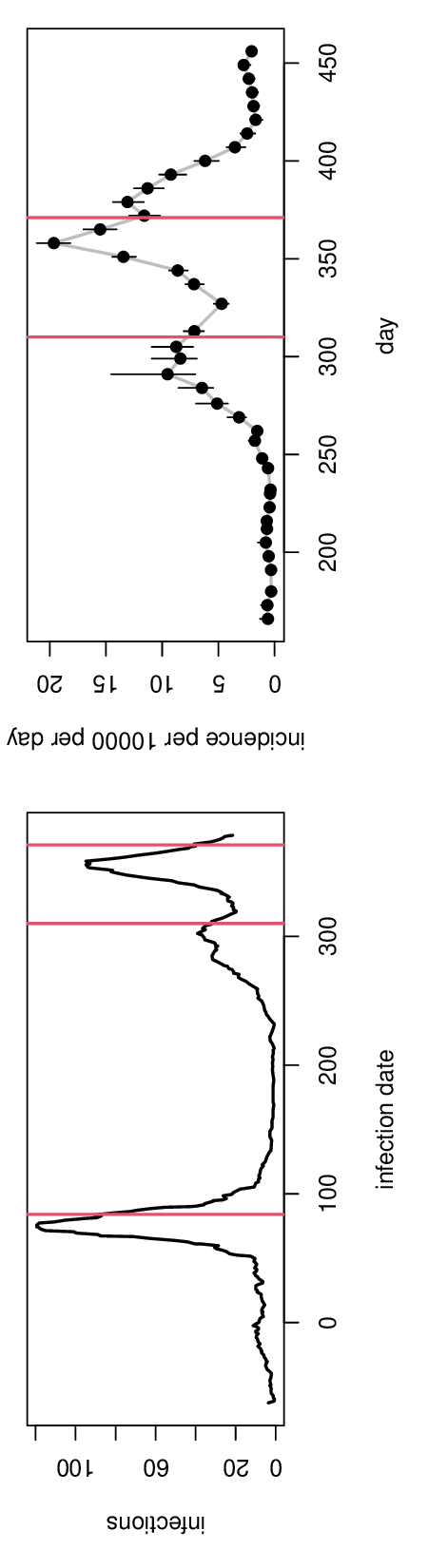}

\vspace*{-.5cm}

\caption{Left, REACT-2 onset of new symptoms per day digitized from \cite{ward2021react}, lagged by the mean 5.8 days from infection to first symptoms. Day 1 is Jan 1 2020. Red lines are the lockdowns. Rather than simple lagging, this paper's method could be applied to the onset dates to infer infection dates given the published  infection to onset distribution, but because none of the onset peaks are after lockdown this can not change the qualitative conclusion (although it would sharpen the peaks a little). Right ONS published estimated incidence and 95\% confidence limits. Red lines are at lockdowns 2 and 3 (the survey was not running at lockdown 1).
\label{ons}}
\end{figure}

The timing result in turn suggests that the measures, and possibly the spontaneous reactions to rising deaths and cases, preceding lockdowns were probably sufficient to bring the epidemic under control, and that community infections, unlike deaths, were probably at a low level well before the first lockdown was eased. Such a scenario would be consistent with the infection profile in Sweden, which began its decline in fatal infections shortly after the UK, but did so on the basis of measures well short of full lockdown. 

The analysis does not in itself say what would have happened without full lockdowns, and must obviously be weighed against other evidence. No currently available analysis will conclusively determine what would have happened without full lockdowns, and the state of the art in causal inference is obviously a very long way from being able to answer this question. Models based on approximations to the mechanisms of epidemic transmission do not allow reliable answers to these causal questions either. This is particularly so given the paucity of data with which to validate their component assumptions - a paucity that only grows more acute as more detail is included in the models. These are not weather or climate models, based on the bulk properties of enormous numbers of physically well understood interactions of simple molecules, tested and refined against huge quantities of carefully measured calibration data collected worldwide over decades. Rather they are best working approximations constructed by experts given the limited information that could be rapidly assembled in a matter of months, and subject to all the uncertainty this implies. A model does not become a valid basis for casual inference merely by being described as mechanistic.  As the above reanalysis using the Flaxman model serves to emphasise: the inclusion of model structure aiming to represent mechanism is no guarantee of improved statistical inference, and certainly not a justification for treating inference with mechanism based models as causal.     

Since this work was first undertaken other low assumption analyses have appeared, in particular looking for the coincidence of NPI introductions and changepoints in incidence, for example in Germany and Spain. The results of this paper are in some alignment with such analyses for Germany \citep{Wieland2020, kuchenhoff2020}, which also suggest that a decline in incidence preceded the first full lockdown. Both are based on case data, which are problematic even in Germany which had mass (but not randomized) testing in place from the start of the epidemic. However it seems likely that the biases in case data would lead to the start of decline in incidence being estimated as later than it really was, rather than earlier, so the qualitative conclusion is likely to be robust. In Spain, \cite{santamaria2020} also identify substantial changes in rate of change of incidence before Spanish lockdown based on death series, but not sufficient to suggest a decline in incidence before lockdown. Based on pre-print versions of the current paper a number of researchers have also attempted to employ the basic idea of dynamic model free inference about incidence profiles, but via a simplified method. This method tries to impute date of infection by subtracting a random draw from the fatal duration distribution from each deceased patient's  death date. This process is replicated to obtain an expected incidence profile. The method is invalid, as duration of disease is not independent of time of death, and it will tend to incorrectly show much less steep, or no, decline before lockdown. See the Supporting Information for a full discussion. 

The results of applying the method to data up to mid February 2021, as well as being consistent with the independent measurements in Figure \ref{ons}, also  provide a picture consistent with the results for the first lockdown. In particular the results preceding the first lockdown appear consistent with how the epidemic progressed under later restrictions short of lockdown. This is not the case for the published analyses suggesting high $R$ and surging incidence on the eve of the first lockdown. The fact that school re-opening does not appear to be followed by an increase in $R$ is interesting: whether it relates to people deciding to keep school children apart from the vulnerable, which is anecdotally plausible, or to other factors, is unclear.  While tempting, it is difficult to interpret the later patterns in terms of the new, apparently more infectious, variant that emerged in late 2020: there is confounding with seasonality of transmission, behavioural changes around the end of year holidays and with the roll out of effective vaccines from late December onwards. Greater clarity on these issues may emerge in future, particularly if the UK ONS Covid surveillance data eventually becomes public in raw form.



\section*{Acknowledgments} 

Thanks to Nicole Augustin, John Coggon, Dan Cookson, Peter Green, Dirk Husmeier, Anna Jewell, Jason Matthiopoulos,  Jonty Rougier and Ernst Wit for various suggestions and discussions of aspects of the paper. 
I am grateful to Robert Verity for his unsolicited offer to help get access to the duration data for England, for making the effort to get the permissions and follow through on this, and for being willing to repeat the original data extraction to remove right truncation problems. I am also very grateful to a non-attributable source who provided key information on the necessity of filtering out patients with onset after hospitalization from such data. I thank two fact checkers who, although they declined to correct an article at {\tt fullfact.org} to remove misrepresentations of this paper, \cite{linton2020incubation} and \cite{flaxman2020lockdown} and some other related statistical flaws, alerted me to the ISARIC study.


The peer review process for this paper was unusual, with 5 referees and 3 editors/associate editors providing comments, which influenced the paper as follows. The paper was first rejected by {\em JRSSC (Applied Statistics)} in early May 2020 for insufficient methodological novelty, and then by {\em Nature} and {\em Nature Communications}, all without refereeing. In late May it was sent to {\em PLOS ONE} and reviewed by an epidemiological referee: they argued that the paper should not be published as, although the substance of the paper appeared correct, a reader might interpret it as implying that lockdowns were unnecessary. They also argued that the drop in incidence could have occurred despite $R>1$ before lockdown: this latter point inspired the simple $R $ calculation presented in the paper. Editor Abdallah M. Samy rejected the paper (at the end of July) on the basis of this first report, a decision formally overturned on appeal in early August. A second referee then strongly recommended publication in a detailed review (dated late August), and provided useful references on the care home epidemic. At the same round a third referee commented briefly on the paper, stating that recessions save lives and consequently suggesting removal of one paragraph of discussion from the paper. On 20 November Abdallah M. Samy formally invited a major revision, but added his own report stating that the paper was unacceptable, arguing that the discussion of possible human costs of lockdown was inappropriate, that incidence {\em can} be meaningfully inferred from case data (how was not stated), that statistical analysis could suggest nothing whatsoever about what might have happened without lockdown (a dynamic epidemic model is apparently required), that the study needed an ethics committee statement, and that the paper was remiss in not citing Flaxman et al (the update including the Flaxman model having been supplied 2 months earlier). These points motivated the transfer of the paper to the statistical journal {\em Biometrics} and the discussion of causal inference in the discussion. One referee and the AE at {\em Biometrics} made the very useful suggestion to analyse the data up to mid February 2021, and provided helpful extra references, substantially improving the paper. The editor's comments were also helpful. A further (theoretical epidemiologist) referee understood the paper's aim to be implementation of the Flaxman et al. model as a GAM, motivating much clearer signposting of several technical aspects. Their comments on age structure effects and the feasibility of PCR testing based incidence estimation led to the sections in the Supplementary Material on these points, and they also suggested giving calendar dates on plots making them much more readable.


\begin{center}
{\Large \bf Supporting Information for `Inferring UK COVID-19 fatal infection trajectories from daily mortality data'}
\end{center}
\setcounter{section}{0}

\section{Feasible direct inference of incidence from randomized PCR testing} 

\begin{figure}[b!]

\eps{-90}{.6}{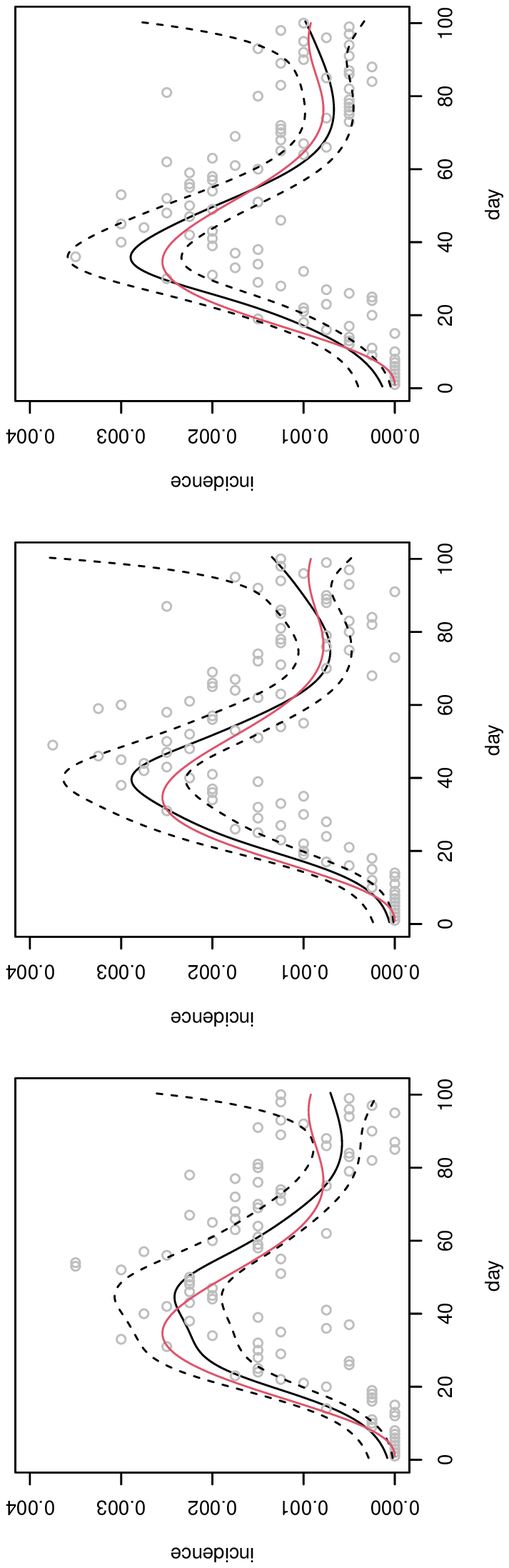}

\caption{Three replicates of incidence rates reconstructed from simulated PCR testing data. True incidence is in red. Reconstructions and 2 standard error bands in black. Grey circles show the number of positive tests each day, divided by 4000. Obviously positivity lags incidence. \label{pcr-fig}}
\end{figure}

Useful estimates of incidence can be obtained from properly randomized PCR surveillance testing, even using numbers of tests well within the laboratory capacity available early in the epidemic. This section provides a simple illustration of this, by sketching a method and showing its ability to capture incidence profiles at the sort of levels that are important for decision making - i.e. at a level slightly over 1 per 1000 per day. For illustrative purposes I consider a very simple model of PCR positivity in which the proportion, $P$, of people potentially testing positive is governed by the simple ODE model ($\dot x$ denoting the time derivative of $x$)
$$
\dot{P} = f(t) - \delta P
$$
where $f$ is the incidence (strictly speaking of potential PCR positivity) as a proportion of the population, and $1/\delta$ is the mean duration of positivity. One could of course substitute any number of alternative models for the assumption of an exponential distribution of the time that subjects are PCR positive, without changing the basic approach discussed here. With only slightly more effort a stochastic formulation could also be substituted (although is likely to add little, given the large numbers involved). The number testing positive in random samples of size $N$ from the population is then given by 
$$
y_i \sim \text{binom}(N,\alpha P)
$$    
where $\alpha $ is the test sensitivity (which is measurable in a reasonably direct manner). As in the main paper, we can represent $f$ semi-parametrically, e.g. using a smoothing spline, so that 
$$
f(t) = \exp ({\bf X}_t {\bm \beta})
$$
where ${\bf X}_t$ is a row vector of spline basis functions evaluated at time $t$. Writing the derivative of $P$ w.r.t. $\beta_j$ as $P_{\beta_j}$ we have an ODE
$$
\dot{P}_{\beta_j} = f(t){ X}_{tj} - \delta P_{\beta_j}
$$
for each such derivative (known as {\em sensitivities} in this context). Given any value of $\bp$ it is straightforward to solve for $P$ and the sensitivities, for example by 4th order Runge-Kutta integration. Hence the log likelihood and its derivative are readily evaluated, and the empirical Bayes approach given in the main paper can be used to find the posterior models, $\hat \bp$, an appropriate smoothing parameter and the large sample posterior covariance matrix. To avoid requiring the second derivative ODE system, $\hat \bp $ can be obtained by quasi-Newton optimization, with the Hessian required for smoothing parameter update obtained from the first derivative of the log likelihood by finite differencing. 

By way of illustration, data were simulated from such a model for 100 days, with 400 tests per day (2800 per week) conducted on randomly selected people from a general population subject to the true incidence curve shown in red in figure \ref{pcr-fig}, and $\delta=0.1$. The method was then used to reconstruct the incidence curve (here 100\% sensitivity was assumed, since sensitivity is a simple scale parameter in this problem). Three random replicate reconstructions are shown in figure \ref{pcr-fig}. Uncertainty is wide at the end of the data, but usable for 10 days earlier. Of course the swab to testing lag adds to this. Larger sample sizes would be needed if local/regional estimates are required, but for the `whole country' picture considered in the main paper such direct estimation is clearly feasible. 

\section{How not to infer fatal incidence}
\begin{figure}
\vspace*{-1cm}

\eps{-90}{.6}{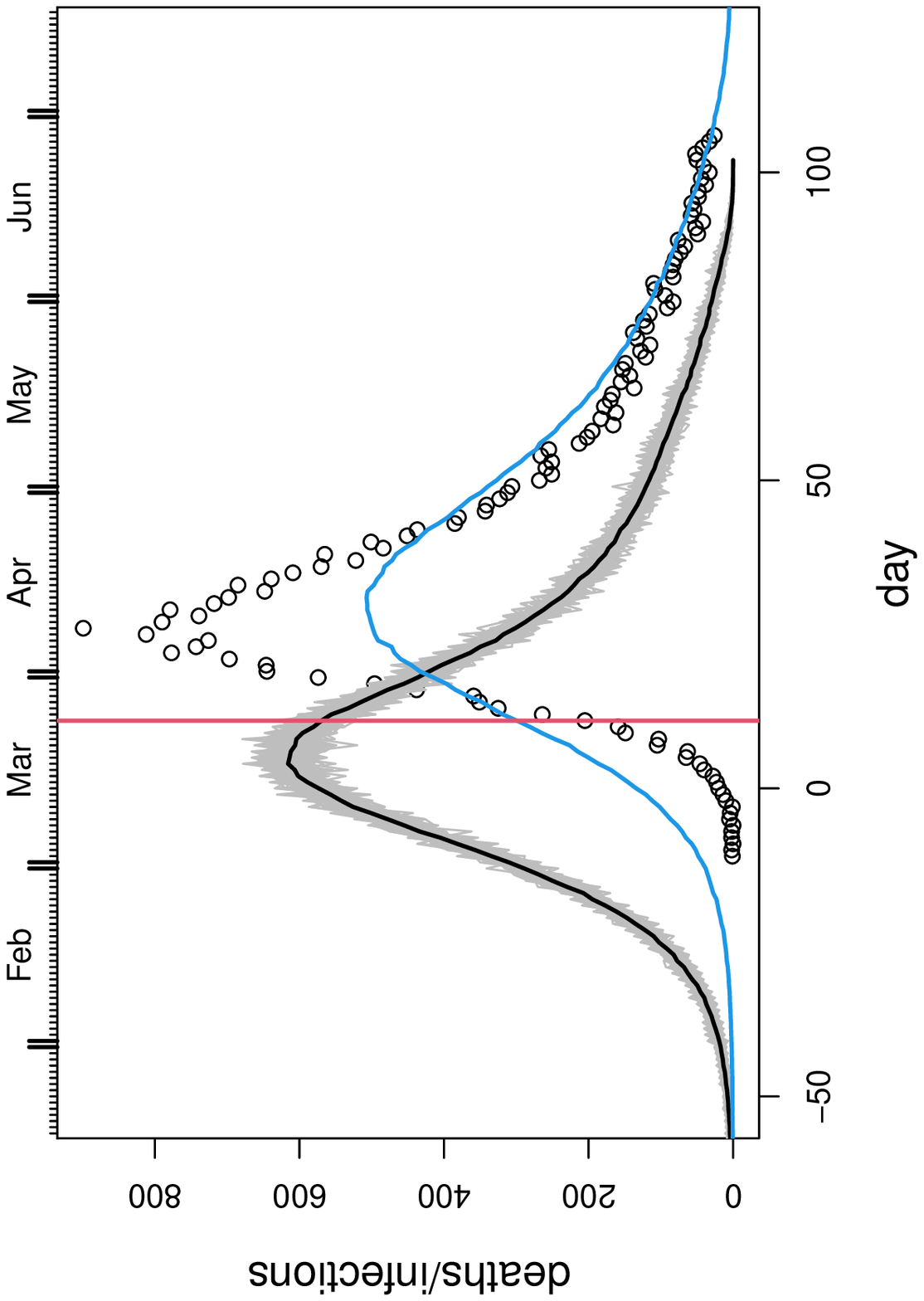}

\caption{Illustration of the failure of a simple imputation method to correctly reconstruct an incidence curve consistent with the observed deaths. Symbols are observed English hospital deaths. Grey curves are infections imputed by the incorrect method given in the text. The black curve is the mean imputed incidence. The blue curve is the expected daily death trajectory implied by the black curve. If the method were correct it should pass through the data points. The red vertical line marks the first day of the first UK lockdown.
\label{wackamole}}
\end{figure}

Several researchers picked up the pre-print version of this paper \citep{wood2020arxiv} and have attempted to use the basic idea of inferring fatal incidence directly from death trajectories and the fatal disease distribution, but via a simple `imputation' method. Suppose the $i$th patient died on day $t_i$. A random draw from the fatal disease duration distribution, $\tau_i$, is subtracted from their death day to give an imputed infection day, $t_i-\tau_i$. Repeating this for all deaths generates an imputed fatal incidence curve. Repeating the imputation many times allows an expected incidence curve to be generated.

This method is not valid. It is completely plausible that duration of disease is independent of time of infection, but not of time of death. Further, unless incidence and deaths are at some constant equilibrium, duration of disease can not be independent of both time of infection {\em and} time of death: when deaths are rising, we inevitably see the deaths from short duration diseases before those from longer durations.  Since the imputation method assumes independence of $t_i$ and $\tau_i$ it can not be valid. Figure \ref{wackamole} shows that this is not a minor concern. It shows incidences reconstructed using the described imputation method. I then added random draws from the fatal duration distribution to the imputed days of infection, to obtain the expected daily deaths implied by the imputed incidence trajectory (essentially the `sanity check' applied in the main paper). The expected daily deaths are an exceedingly poor fit to the data.

\section{Fatal disease duration distribution}

Fatal disease duration data for England are available in the CHESS\footnote{COVID-19 Hospitalisations in England Surveillance System} database, access to which is restricted to particular research groups under strict conditions. With the kind help of Robert Verity from Imperial College I was able to access information on the distribution of fatal disease durations for 3274 deaths that occurred before 10 June 2020 with recorded symptom onset before 1 May. The information provided was a bar chart of the duration distribution by day, on condition that only the information about the model fitted to the data be distributed further. The data were not filtered to remove hospital acquired infections, but it was not possible to obtain data only for those with onset before hospitalization. This is problematic for two reasons. Firstly, for inferring the time course of community acquired fatal infections it is the distribution of fatal disease durations for community acquired infections that is required, which the raw data do not provide: for example, they contain substantial proportions of durations of 1-3 days that appear clinically implausible for deaths from community acquired COVID-19 \citep[see, e.g.][]{huang2020clinical,wang2020clinical,zhou2020clinical, tay2020trinity}. Secondly the raw data are from  a relatively small proportion of the total deaths. It is very unlikely that the ratio of hospital to community acquired infections in this sample is representative: for hospital acquired infections the onset of symptoms is presumably almost always known, and hence more likely to be recorded than for community acquired infections. This makes the raw distribution unrepresentative of the distribution for all deaths and also not usefully informative about the proportion of all deaths that are from hospital acquired infection. Note also that without more extensive data access it is not possible to rule out that some proportion of what appear to be hospital acquired infections really represent data problems (for example recording onset day as hospital admission day).   

\begin{figure}

\vspace*{-1cm}

\eps{-90}{.7}{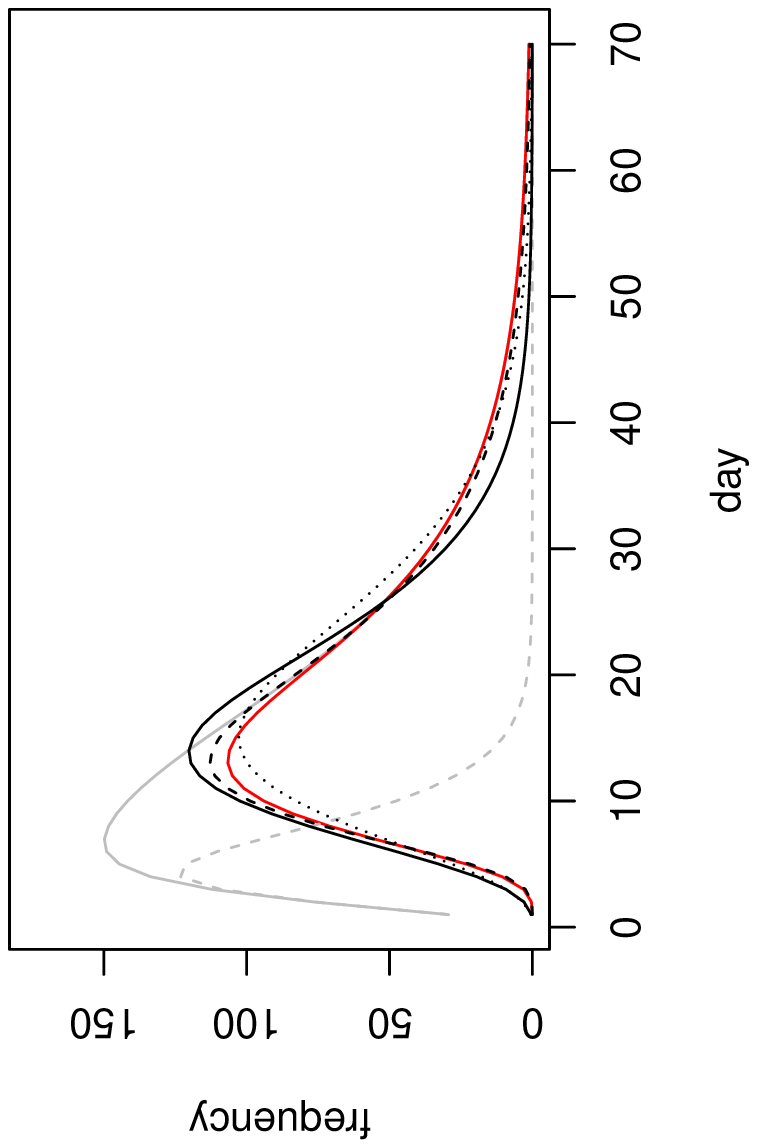}

\vspace*{-.5cm}

\caption{Onset to death duration distribution models. The red curve is the log-normal mixture component for community acquired infection fitted to the CHESS data, the dashed grey curve is the gamma mixture component representing hospital acquired infection and the continuous grey curve the combined model. The combined model is not directly usable: see text. The black curves are: continuous \cite{verity2020ifr}; dashed \cite{linton2020incubation}; dotted \cite{wu2020covid}. The mixture model was estimated by maximum likelihood, with the hospital acquired mixture proportion reduced until the profiled log likelihood was reduced to 4 below the MLE, to obtain the shortest mean community acquired duration consistent with the data under this model. The black curves in no way inform the red curve in the fitting.  \label{o2ds}}
\end{figure}

To deal with these issues a two component mixture model was fitted to data digitized from the bar chart, consisting of a gamma distribution (representing hospital acquired infections) and a log-normal distribution (representing community acquired infections). Parameterization was such that the log-normal had the longer mean duration. The higher the gamma mixture proportion the larger the log-normal mean. To find the shortest mean community acquired duration defensible from the data, the gamma mixture proportion was reduced to the point at which the log likelihood was about 4 below the MLE (decreasing further decreases the log-likelihood sharply, pushes a $\chi^2$ goodness of fit statistic into the significant range, and starts to suggest rather high probabilities of very short disease durations for the log-normal mixture component). This point has about 0.7 of the mixture contributed by the community infection component. The resulting log-normal community infection fit has a mean of 21 days and a standard deviation of 12.7. Longer durations would be slightly more consistent with the data under the mixture model, but given the aims of this paper it is better to use conservatively short estimates here. Figure \ref{o2ds} shows the various estimated distributions over the duration range observed in the CHESS data. The log-normal model has an earlier mode, but longer tail, than the \cite{verity2020ifr} model used in earlier versions of this paper.

It should perhaps be noted that this model was obtained before I was aware of \cite{linton2020incubation} and  \cite{wu2020covid}. Note also that the data for this model were obtained before the decision to attribute deaths to Covid-19 only if there was a positive test within the 28 days preceding death: this may be the reason for the model's slightly heavier tail. Otherwise the results are broadly in agreement with those from the published studies.

Assuming independence of incubation period and onset to death period, the preceding fit and the \cite{McAloon20} incubation period imply that the infection-to-death distribution for the community acquired infection component can be well modelled by a log-normal distribution with log scale mean and standard deviation of 3.19 and 0.44, respectively. That is a mean of 26.8 days and standard deviation of 12.4 days. The community infection distribution component is shown in blue in figure 2 of the paper.

More recently results of Robert Verity's own more detailed analysis of the CHESS data have appeared in \cite{rep41}. The full fitted distribution is not given, but the figures that are reported imply a slightly shorter mean duration of just over 24 days. This is just under a day and  a half less than for the mean duration for the average distribution used in the main paper, and within the uncertainty range considered in the paper.  

\begin{figure}


\eps{-90}{.6}{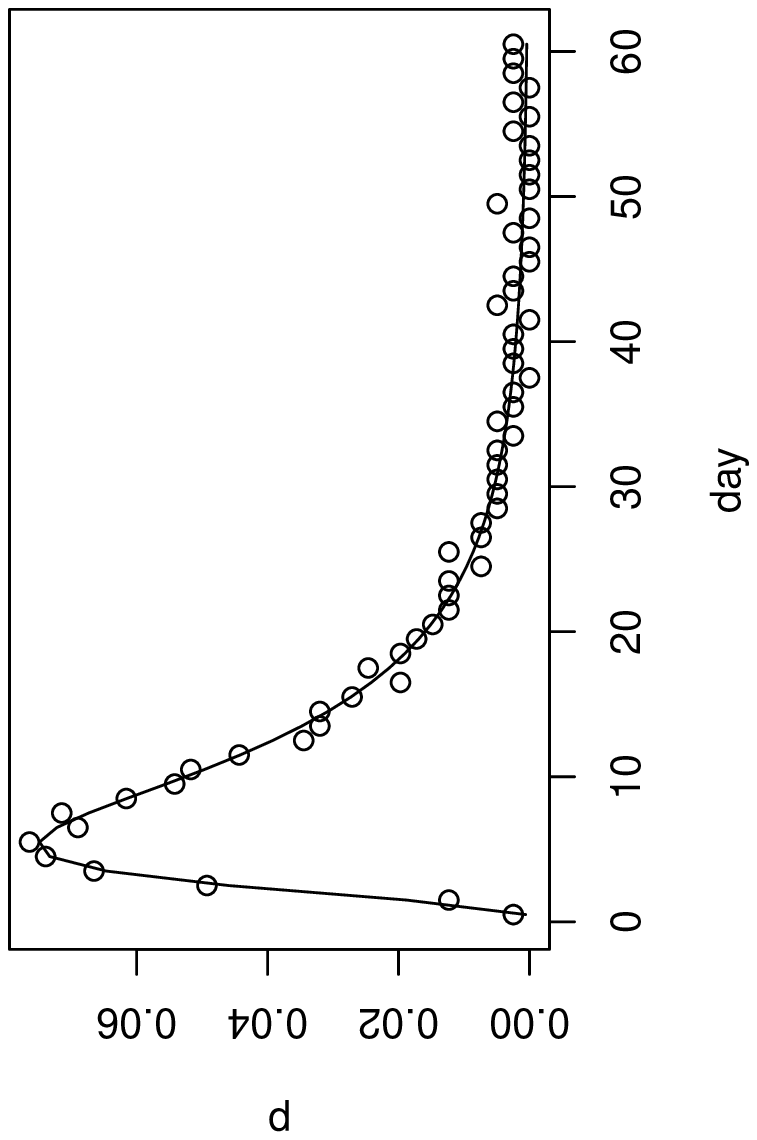}

\vspace*{-.5cm}

\caption{Empirical probability function of hospitalization to death interval distribution obtained by differencing the CDF digitized from Figure 12 of \cite{isaric.oct20}, fitted (to minimize KL-divergence) by a log normal density.  \label{isaric-pdf}}
\end{figure}

A second less problematic source of information is the ISARIC study, which I was unaware of before the peer reviewed version of this paper was published. \cite{isaric.oct20} summarizes information on the patients enrolled in this study up until October 2020. At this time point the recruitment rate to the study was quite low, so that right truncation problems should have a fairly low impact. Figure 12 of the study plots the observed cumulative distribution function for time from hospitalization to death for the 24421 study patients who had died up to the time of the report. This can be digitized and converted to a probability function, which can be well approximated by a log-normal density (mean 12.47 and standard deviation 10.97) - see Figure \ref{isaric-pdf}. \cite{isaric.oct20} report that the time from onset of symptoms to hospitalization had a mean of 7.7 days and standard deviation of 6.1. So the mean time from onset to death is about 20.2 days. Assuming independence of the two durations the standard deviation of the onset to death duration is then 12.55.

Modelling the onset to hospital duration as log normal, the best fit lognormal approximation to the distribution of time from infection to death then has a mean of 27.7 and standard deviation 12.0 (an alternative gamma approximation was a poor fit). Note that this distribution estimate does not take into account the small amount of right truncation present in the data, so might be slightly biased towards lower durations.

\section{Possible age structure effects}

One possible concern is that if the distribution of fatal disease duration is strongly age dependent and the age distribution shifts over time, then the results of the paper's analysis could be biased in ways that could be difficult to correct. In fact \cite{dennis2021covid-improve} looked for temporal changes in patient characteristics including age as possible explanations for the mortality improvements that they report in the early months of the epidemic, but did not find age distribution changes in hospitalized patients. Additionally \cite{rep41} analysed English hospital data to parameterize a detailed age structured epidemic and hospital model, but while they report age effects on rates of hospitalization and transfer to ICU, with different distributions of time to death for ICU and general ward patients, those distributions are not reported to be age dependent. 

\begin{figure}


\eps{-90}{.6}{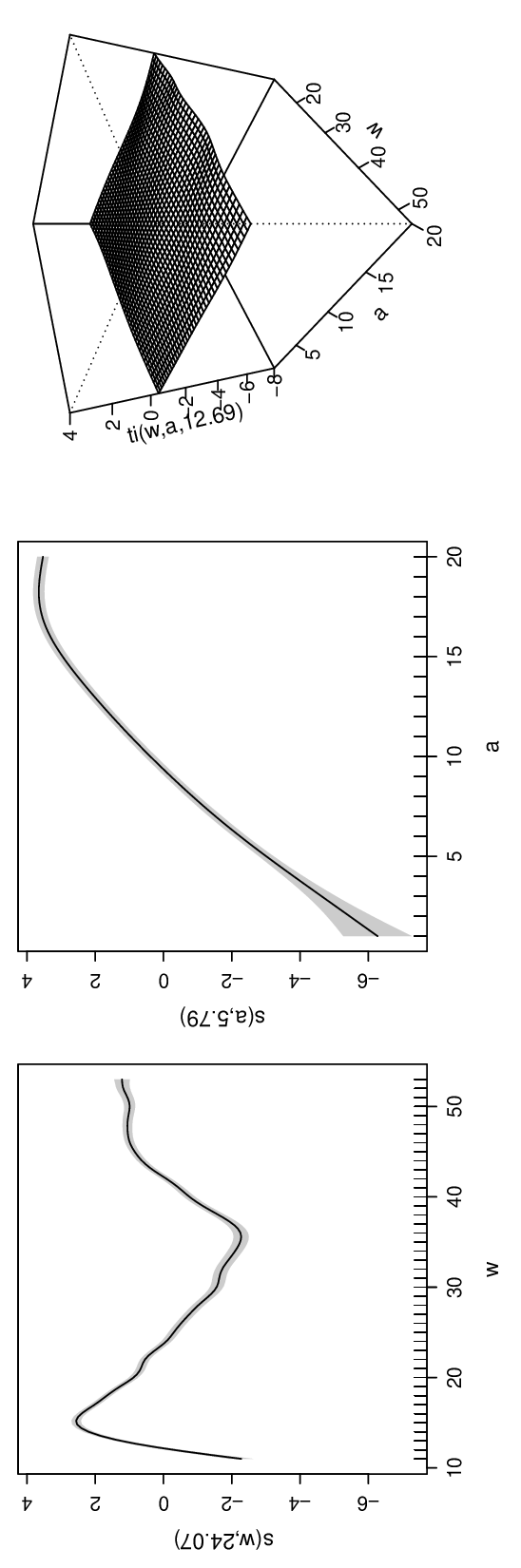}

\vspace*{-.5cm}

\caption{Generalized additive model term estimates on the log link scale, from fitting to England and Wales hospital death by age data. Left is the effect of week, middle of age group and right of their interaction. The interaction ranges over approximately -0.2 to 0.2, and is clearly a very small effect relative to the others.  \label{age-shift}}
\end{figure}

While both the cited analyses rely on confidential data with stringently controlled access, it is possible to look for evidence of age distribution shifts in the weekly England and Wales Covid-19 deaths by age data publicly available from the UK Office for National Statistics\footnote{{\tiny https://www.ons.gov.uk/peoplepopulationandcommunity/birthsdeathsandmarriages/deaths/datasets/weeklyprovisionalfiguresondeathsregisteredinenglandandwales}}. These data give total England and Wales Covid-19 deaths each week in 20 age bands, $<1$, 1-4, 5-9 , \ldots, 85-89 and 90+. They also record the total number of Covid-19 deaths each week in care homes for the elderly in England and Wales. To look for age distribution changes in hospitalized patients, it is necessary to remove the care home deaths from the weekly totals. The care home deaths are not broken down by age, so I simply reduced the total deaths in the last three age classes by the same proportion, in order to reduce the each weekly total deaths by the correct amount. 

A negative binomial generalized additive model was then fitted to the data, with the structure
$$
\log\{E({\tt death}_i)\} = \alpha + f_1(a_i) + f_2(w_i) + f_3(a_i,w_i)
$$
where $a_i$ denotes the age class (a number from 1 to 20) and $w_i$ is the week. $f_1$ and $f_2$ are univariate splines, while $f_3$ is a tensor product interaction spline, (without the main effects). Thus $f_3$ represents any change in age distribution of deaths over time. See \cite{wood2017igam} section 5.6.3 for details. $f_3$ is statistically significant, but the effect size is too small to be biologically significant. 
Figure \ref{age-shift} shows the estimated model components. Leaving the care home deaths in the totals leads to a slightly stronger interaction ranging from about -0.6 to 0.4 in the early weeks. This reflects the somewhat different dynamics of the care home epidemic relative to the community epidemic, as discussed in the main paper. The main effects are essentially unchanged from those shown.

\section{Further model checking of relaxed Flaxman model}

\begin{figure}
\vspace*{-.2cm}  
\eps{-90}{.5}{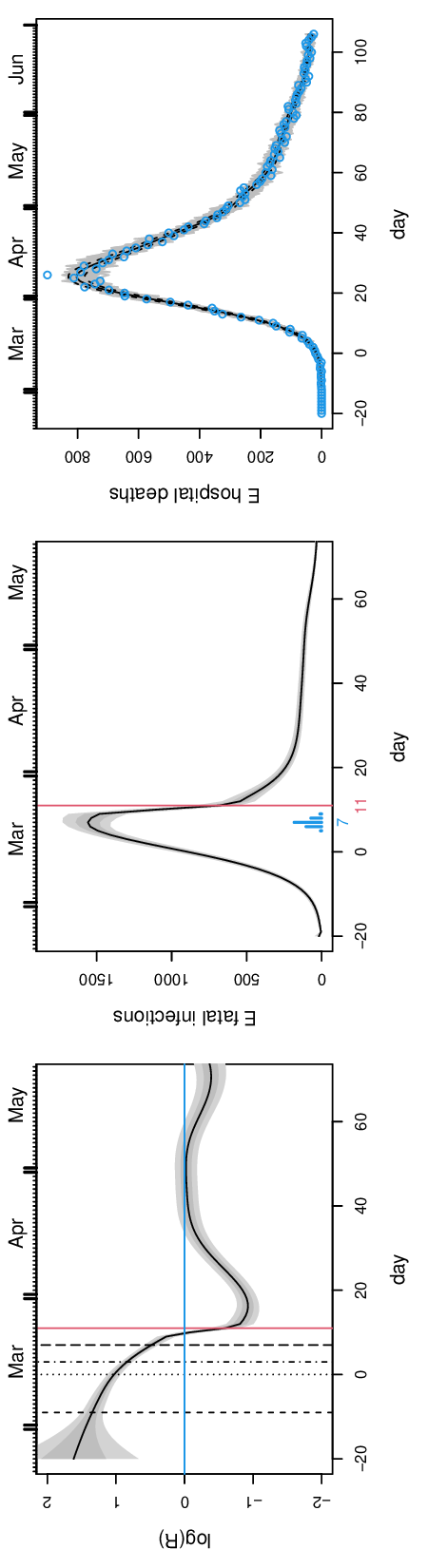}
\vspace*{-.3cm} 

\eps{-90}{.5}{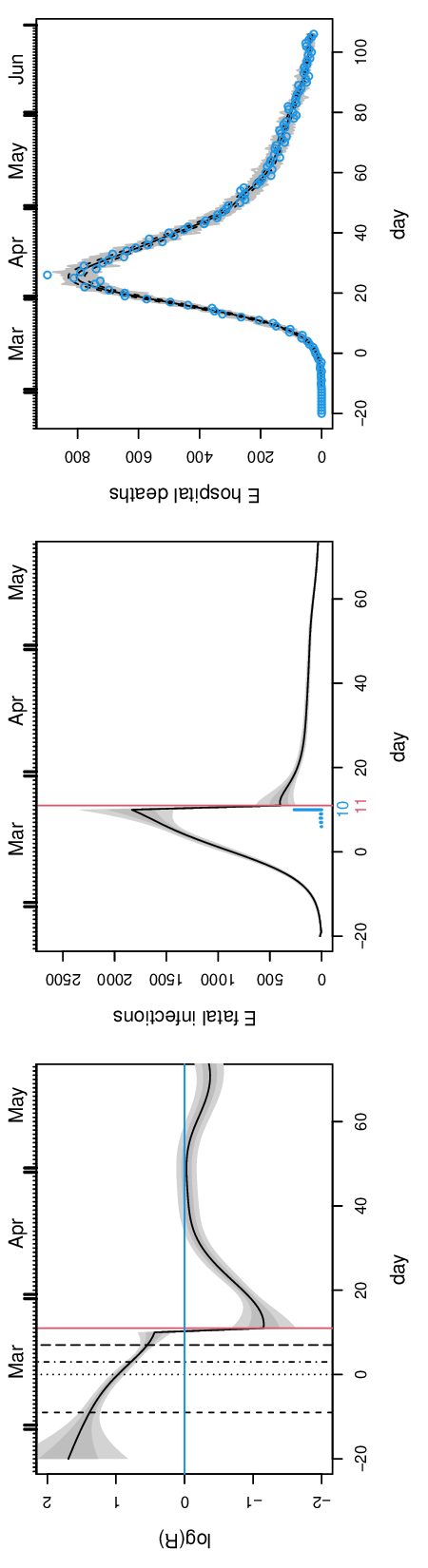}

\vspace*{-.2cm} 

\caption{Model checking plots for the \cite{flaxman2020lockdown} model. The upper row shows the results of applying a time dilation around lockdown to ensure that any very rapid change in $R$ at that point can be accommodated by the model. The results are similar to the undilated case. The lower row shows a model which forces a step change at lockdown- notice the severe boundary uncertainty in the vicinity of lockdown introduced by this (see text for discussion). Even with this model $R$ is about 1.5, substantially below the \cite{flaxman2020lockdown} estimates of around 3 on the eve of lockdown. \label{flaxman-check}
}
\end{figure}  

The time dilation check from the {\em Model checking} section of the paper was also applied to the relaxed \cite{flaxman2020lockdown} model, with the results shown in the upper panel of figure \ref{flaxman-check}. Again the results are qualitatively similar to the undilated case, despite modifying the model to favour sharp change in $R$ at lockdown. Although highly problematic for the reasons discussed in the paper, the results of a check using a model in which a step change was forced to occur at lockdown is also shown in the lower row of figure \ref{flaxman-check}. The boundary condition artefacts that this introduces are clearly visible, but in any case the inferred $R$ on the eve of lockdown is about 1.5. This is substantially below the Flaxman et al. estimates of close to 3.  
    
\section{Sensitivity to mortality rate reductions}

\begin{figure}
 
\eps{-90}{.5}{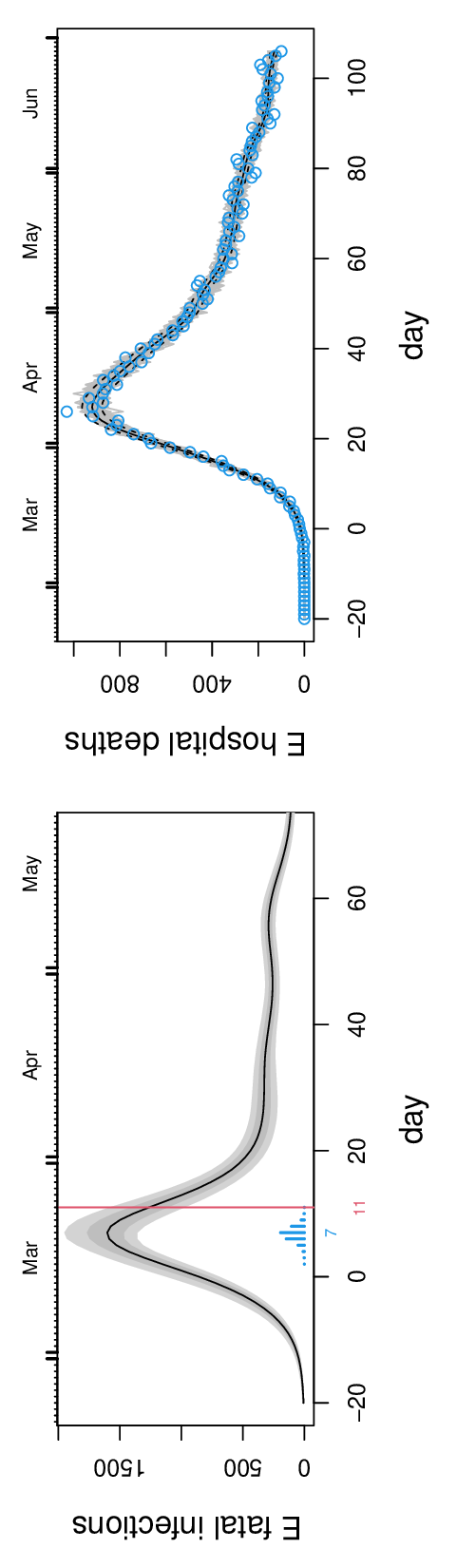}

\caption{Sensitivity of the results to improvements in the IFR. To interpret the fatal incidence trajectories as proportional to overall disease incidence, the IFR has to be constant. There is evidence for this not being the case as hospital care has improved. These plots show inferred incidence from death data `corrected' for the mortality rate improvements estimated in \cite{dennis2021covid-improve}. Note the very slight rightward shift in the peak timing distribution, and somewhat slower decay in the incidence profile. \label{ifr-check}
}
\end{figure}  

There is evidence for reductions in the hospital mortality rates in England from the week of 29th March 2020 until the end of June, with this reduction apparently not being attributable to any change in patient characteristics: \cite{dennis2021covid-improve} report mortality rates reducing by a multiplicative factor of about .985 per day (before then, if anything the death rates were increasing). While this does not undermine inference of fatal infection incidence, it obviously means that fatal disease incidence should probably not be interpreted as proportional to overall incidence. Given the uncertainties in the \cite{dennis2021covid-improve} results, a direct correction is difficult. Furthermore ruling out changes in severity of disease required for  admission over the first wave is also not possible: for example, general practitioners (family doctors) were initially working with central guidance on when patients should self isolate, but not when they should be sent to hospital, so it seems unlikely that on the ground admission criteria were constant, especially at times when some hospitals were at or near capacity. However a sensitivity test is straightforward. The observed deaths each day can simply be scaled up by the ratio of the number of deaths expected without improvements to the number expected with improvements (assuming .985 per day improvement from 29th March). This has the effect of making the downward tail of the adjusted deaths series decay more slowly than for the observed deaths (see right panel of fig \ref{ifr-check}). Applying the method to the English hospital data then gives the results in figure \ref{ifr-check}. There is a shift in the inferred peak incidence to later, and the incidence decays more slowly, relative to the results shown in the main paper. Note that the mortality improvements only apply to hospital deaths, not care home deaths.


\begin{thebibliography}{}

\bibitem[\protect\citeauthoryear{Anon}{Anon}{2020}]{covid-life-loss-govt}
Anon (2020).
\newblock Direct and indirect impacts of {COVID-19} on excess deaths and
  morbidity.
\newblock Technical report, Department of Health and Social Care, Office for
  National Statistics, Government Actuary’s Department and Home Office.

\bibitem[\protect\citeauthoryear{Comas-Herrera, Zalaka{\'\i}n, Litwin, Hsu,
  Lane, and Fern{\'a}ndez}{Comas-Herrera et~al.}{2020}]{comas2020care-home}
Comas-Herrera, A., Zalaka{\'\i}n, J., Litwin, C., Hsu, A.~T., Lane, N., and
  Fern{\'a}ndez, J.-L. (2020).
\newblock Mortality associated with {COVID}-19 outbreaks in care homes: early
  international evidence.
\newblock {\em LTCcovid. org, International Long-Term Care Policy Network} .

\bibitem[\protect\citeauthoryear{Dennis, McGovern, Vollmer, and Mateen}{Dennis
  et~al.}{2021}]{dennis2021covid-improve}
Dennis, J.~M., McGovern, A.~P., Vollmer, S.~J., and Mateen, B.~A. (2021).
\newblock Improving survival of critical care patients with coronavirus disease
  2019 in england: a national cohort study, {M}arch to {J}une 2020.
\newblock {\em Critical care medicine} {\bf 49,} 209.

\bibitem[\protect\citeauthoryear{Diekmann, Heesterbeek, and Metz}{Diekmann
  et~al.}{1990}]{diekmann1990R}
Diekmann, O., Heesterbeek, J. A.~P., and Metz, J.~A. (1990).
\newblock On the definition and the computation of the basic reproduction ratio
  {R0} in models for infectious diseases in heterogeneous populations.
\newblock {\em Journal of mathematical biology} {\bf 28,} 365--382.

\bibitem[\protect\citeauthoryear{Domanski, Scatigna, and Zabai}{Domanski
  et~al.}{2016}]{domanski2016}
Domanski, D., Scatigna, M., and Zabai, A. (2016).
\newblock Wealth inequality and monetary policy.
\newblock {\em BIS Quarterly Review March} .

\bibitem[\protect\citeauthoryear{Flaxman, Mishra, Gandy, Unwin, Mellan,
  Coupland, Whittaker, Zhu, Berah, Eaton, et~al\mbox{.}}{Flaxman
  et~al.}{2020}]{flaxman2020lockdown}
Flaxman, S., Mishra, S., Gandy, A., Unwin, H. J.~T., Mellan, T.~A., Coupland,
  H., et~al. (2020).
\newblock Estimating the effects of non-pharmaceutical interventions on
  {COVID-19} in {E}urope.
\newblock {\em Nature} {\bf 584,} 257--261.

\bibitem[\protect\citeauthoryear{Folkh\"alsomyndigheten}{Folkh\"alsomyndigheten}{2020}]{sweden-covid-daily}
Folkh\"alsomyndigheten (2020).
 experience.arcgis.com/\\experience/09f821667ce64bf7be6f9f87457ed9aa.

\bibitem[\protect\citeauthoryear{Fontan, Claveau, and Dietsch}{Fontan
  et~al.}{2016}]{fontan2016QE}
Fontan, C., Claveau, F., and Dietsch, P. (2016).
\newblock Central banking and inequalities: Taking off the blinders.
\newblock {\em Politics, Philosophy \& Economics} {\bf 15,} 319--357.

\bibitem[\protect\citeauthoryear{Green and Silverman}{Green and
  Silverman}{1994}]{green.silverman}
Green, P.~J. and Silverman, B.~W. (1994).
\newblock {\em Nonparametric Regression and Generalized Linear Models}.
\newblock Chapman \& Hall.

\bibitem[\protect\citeauthoryear{Hanlon, Chadwick, Shah, Wood, Minton,
  McCartney, Fischbacher, Mair, Husmeier, and Matthiopoulos}{Hanlon
  et~al.}{2020}]{hanlon2020}
Hanlon, P., Chadwick, F., Shah, A., Wood, R., Minton, J., McCartney, G.,
  Fischbacher, C., Mair, F.~S., Husmeier, D., and Matthiopoulos, J. (2020).
\newblock {COVID}-19 – exploring the implications of long-term condition type
  and extent of multimorbidity on years of life lost: a modelling study.
\newblock https://wellcomeopenresearch.org/articles/5-75.

\bibitem[\protect\citeauthoryear{Knock, Whittles, Lees, Perez~Guzman, Verity,
  Fitzjohn, Gaythorpe, Imai, Hinsley, Okell, Rosello, Kantas, Walters, Bhatia,
  Watson, Whittaker, Cattarino, Boonyasiri, Djaafara, Fraser, Fu, Wang, Xi,
  Donnelly, Jauneijaite, Laydon, White, Ghani, Ferguson, Cori, and
  Baguelin}{Knock et~al.}{2020}]{rep41}
Knock, E.~S., Whittles, L.~K., Lees, J.~A., Perez~Guzman, P.~N., Verity, R.,
  Fitzjohn, R.~G., et~al.
  (2020).
\newblock Report 41: The {2020 SARS-CoV-2} epidemic in {E}ngland: key
  epidemiological drivers and impact of interventions.
\newblock {\em Imperial College London} .

\bibitem[\protect\citeauthoryear{K{\"u}chenhoff, Guenther, H{\"o}hle, and
  Bender}{K{\"u}chenhoff et~al.}{2020}]{kuchenhoff2020}
K{\"u}chenhoff, H., Guenther, F., H{\"o}hle, M., and Bender, A. (2020).
\newblock Analysis of the early {C}ovid-19 epidemic curve in {G}ermany by
  regression models with change points.
\newblock {\em medRxiv} .

\bibitem[\protect\citeauthoryear{Lauer, Grantz, Bi, Jones, Zheng, Meredith,
  Azman, Reich, and Lessler}{Lauer et~al.}{2020}]{lauer2020covid}
Lauer, S.~A., Grantz, K.~H., Bi, Q., Jones, F.~K., Zheng, Q., Meredith, H.~R.,
 et~al. (2020).
\newblock The incubation period of coronavirus disease 2019 ({COVID}-19) from
  publicly reported confirmed cases: estimation and application.
\newblock {\em Annals of internal medicine} {\bf 172,} 577--582.

\bibitem[\protect\citeauthoryear{Levine, Ramsay, and Smidt}{Levine
  et~al.}{2001}]{levine2001}
Levine, D.~M., Ramsay, P.~P., and Smidt, R.~K. (2001).
\newblock {\em Applied {S}tatistics for {E}ngineers and {S}cientists}.
\newblock Prentice Hall.

\bibitem[\protect\citeauthoryear{Linton, Kobayashi, Yang, Hayashi,
  Akhmetzhanov, Jung, Yuan, Kinoshita, and Nishiura}{Linton
  et~al.}{2020}]{linton2020incubation}
Linton, N.~M., Kobayashi, T., Yang, Y., Hayashi, K., Akhmetzhanov, A.~R., Jung,
  S.-m., et~al. (2020).
\newblock Incubation period and other epidemiological characteristics of 2019
  novel coronavirus infections with right truncation: a statistical analysis of
  publicly available case data.
\newblock {\em Journal of clinical medicine} {\bf 9,} 538.

\bibitem[\protect\citeauthoryear{Marmot, Allen, Boyce, Goldblatt, and
  Morrison}{Marmot et~al.}{2020}]{marmot-review-10}
Marmot, M., Allen, J., Boyce, T., Goldblatt, P., and Morrison, J. (2020).
\newblock {\em Health {E}quity in {E}ngland: {T}he {M}armot {R}eview 10 {Y}ears
  {O}n}.
\newblock The {H}ealth {F}oundation.

\bibitem[\protect\citeauthoryear{McAloon, Collins, Hunt, Barber, Byrne, Butler,
  Casey, Griffin, Lane, McEvoy, et~al\mbox{.}}{McAloon
  et~al.}{2020}]{McAloon20}
McAloon, C., Collins, {\'A}., Hunt, K., Barber, A., Byrne, A.~W., Butler, F.,
  et~al. (2020).
\newblock Incubation period of {COVID}-19: a rapid systematic review and
  meta-analysis of observational research.
\newblock {\em BMJ open} {\bf 10,} e039652.

\bibitem[\protect\citeauthoryear{NHS}{NHS}{2020}]{nhs-covid-daily}
NHS (2020).
\newblock  www.england.nhs.uk/statistics/statistical-work-areas/covid-19-daily-deaths/.

\bibitem[\protect\citeauthoryear{OBR}{OBR}{2020}]{obrjuly20}
OBR (2020).
\newblock Office for budget responsibility fiscal sustainability report {J}uly
  2020.

\bibitem[\protect\citeauthoryear{{Office for National Statistics}}{{Office for
  National Statistics}}{2019}]{ons-lifetables}
{Office for National Statistics} (2019).
\newblock National life tables, {UK}: 2016 to 2018.

\bibitem[\protect\citeauthoryear{{Office for National Statistics}}{{Office for
  National Statistics}}{2020}]{ons-deaths}
{Office for National Statistics} (2020).
\newblock Deaths registered weekly in {E}ngland and {W}ales.

\bibitem[\protect\citeauthoryear{Plummer}{Plummer}{2003}]{plummer2003jags}
Plummer, M. (2003).
\newblock {JAGS}: A program for analysis of {B}ayesian graphical models using
  {G}ibbs sampling.
\newblock In {\em Proceedings of the 3rd International Workshop on Distributed
  Statistical Computing (DSC 2003).}, pages 20--22.

\bibitem[\protect\citeauthoryear{Plummer, Best, Cowles, and Vines}{Plummer
  et~al.}{2006}]{coda}
Plummer, M., Best, N., Cowles, K., and Vines, K. (2006).
\newblock coda: Convergence diagnosis and output analysis for {MCMC}.
\newblock {\em R News} {\bf 6,} 7--11.

\bibitem[\protect\citeauthoryear{Pritchard, Dankwa, Hall, Baillie, Carson,
  Docherty, Donnelly, Dunning, Fraser, Hardwick, et~al.}{Pritchard
  et~al.}{2020}]{isaric.oct20}
Pritchard, M., E.~A. Dankwa, M.~Hall, J.~K. Baillie, G.~Carson, A.~Docherty,
  C.~A. Donnelly, J.~Dunning, C.~Fraser, H.~Hardwick, et~al. (2020).
\newblock {ISARIC} clinical data report 4 {O}ctober 2020.
\newblock {\em medRxiv\/}.

\bibitem[\protect\citeauthoryear{Santamar\'ia and Hortal}{Santamar\'ia and
  Hortal}{2020}]{santamaria2020}
Santamar\'ia, L. and Hortal, J. (2020).
\newblock Chasing the ghost of infection past: identifying thresholds of change
  during the {COVID}-19 infection in {S}pain.
\newblock {\em Epidemiology \& Infection} pages 1--19.

\bibitem[\protect\citeauthoryear{Stoner, Economou, and Halliday}{Stoner
  et~al.}{2020}]{stoner2020deathcasting}
Stoner, O., Economou, T., and Halliday, A. (2020).
\newblock A {P}owerful {M}odelling {F}ramework for {N}owcasting and
  {F}orecasting {COVID}-19 and {O}ther {D}iseases.
\newblock {\em arXiv preprint arXiv:1912.05965} .

\bibitem[\protect\citeauthoryear{Verity, Okell, Dorigatti, Winskill, Whittaker,
  Imai, Cuomo-Dannenburg, Thompson, Walker, Fu, et~al\mbox{.}}{Verity
  et~al.}{2020}]{verity2020ifr}
Verity, R., Okell, L.~C., Dorigatti, I., Winskill, P., Whittaker, C., Imai, N.,
   et~al. (2020).
\newblock Estimates of the severity of coronavirus disease 2019: a model-based
  analysis.
\newblock {\em The Lancet Infectious Diseases} {\bf 20,} 669--677.

\bibitem[\protect\citeauthoryear{Walker, Whittaker, Watson, Baguelin, Ainslie,
  Bhatia, Bhatt, Boonyasiri, Boyd, Cattarino, et~al\mbox{.}}{Walker
  et~al.}{2020}]{report12}
Walker, P., Whittaker, C., Watson, O., Baguelin, M., Ainslie, K., Bhatia, S.,
  et~al. (2020).
\newblock Report 12: The global impact of {COVID-19} and strategies for
  mitigation and suppression.
\newblock {\em Imperial {C}ollege {L}ondon} .

\bibitem[\protect\citeauthoryear{Ward, Cooke, Whitaker, Redd, Eales, Brown,
  Collet, Cooper, Daunt, Jones, et~al.}{Ward et~al.}{2021}]{ward2021react}
Ward, H., G.~Cooke, M.~Whitaker, R.~Redd, O.~Eales, J.~C. Brown, K.~Collet,
  E.~Cooper, A.~Daunt, K.~Jones, et~al. (2021).
\newblock React-2 round 5: increasing prevalence of sars-cov-2 antibodies
  demonstrate impact of the second wave and of vaccine roll-out in england.
\newblock {\em medRxiv\/}.

\bibitem[\protect\citeauthoryear{Wieland}{Wieland}{2020}]{Wieland2020}
Wieland, T. (2020).
\newblock A phenomenological approach to assessing the effectiveness
  of{COVID}-19 related nonpharmaceutical interventions in {G}ermany.
\newblock {\em Safety Science} {\bf 131,} 104924.

\bibitem[\protect\citeauthoryear{Wood}{Wood}{2017}]{wood2017igam}
Wood, S.~N. (2017).
\newblock {\em Generalized Additive Models: An Introduction with R}.
\newblock CRC press, Boca Raton, FL, 2 edition.

\bibitem[\protect\citeauthoryear{Wood and Fasiolo}{Wood and
  Fasiolo}{2017}]{wood2016gfs}
Wood, S.~N. and Fasiolo, M. (2017).
\newblock A generalized {F}ellner-{S}chall method for smoothing parameter
  optimization with application to {T}weedie location, scale and shape models.
\newblock {\em Biometrics} {\bf 73,} 1071--1081.

\bibitem[\protect\citeauthoryear{Wood, Pya, and S{\"a}fken}{Wood
  et~al.}{2016}]{wood2015plig}
Wood, S.~N., Pya, N., and S{\"a}fken, B. (2016).
\newblock Smoothing parameter and model selection for general smooth models
  (with discussion).
\newblock {\em Journal of the American Statistical Association} {\bf 111,}
  1548--1575.

\bibitem[\protect\citeauthoryear{Wood and Wit}{Wood and
  Wit}{2021}]{wood2021rep41}
Wood, S.~N. and Wit, E.~C. (2021).
\newblock Was ${R}< 1$ before the {E}nglish lockdowns? {O}n modelling
  mechanistic detail, causality and inference about {C}ovid-19.
\newblock {\em medRxiv} .

\bibitem[\protect\citeauthoryear{Wood, Wit, Fasiolo, and Green}{Wood
  et~al.}{2020}]{lancet-ifr}
Wood, S.~N., Wit, E.~C., Fasiolo, M., and Green, P.~J. (2020).
\newblock {COVID}-19 and the difficulty of inferring epidemiological parameters
  from clinical data.
\newblock {\em The Lancet Infectious Diseases} .

\bibitem[\protect\citeauthoryear{Worldometer}{Worldometer}{2020}]{worldometer}
Worldometer (2020).
\newblock www.worldometers.info/coronavirus/.

\bibitem[\protect\citeauthoryear{Wu, Leung, Bushman, Kishore, Niehus,
  de~Salazar, Cowling, Lipsitch, and Leung}{Wu et~al.}{2020}]{wu2020covid}
Wu, J.~T., Leung, K., Bushman, M., Kishore, N., Niehus, R., de~Salazar, P.~M.,
 et~al. (2020).
\newblock Estimating clinical severity of {COVID}-19 from the transmission
  dynamics in {W}uhan, {C}hina.
\newblock {\em Nature Medicine} {\bf 26,} 506--510.

\end{thebibliography}

\begin{thebibliography}{}

\bibitem[\protect\citeauthoryear{Dennis, McGovern, Vollmer, and Mateen}{Dennis
  et~al.}{2021}]{dennis2021covid-improve}
Dennis, J.~M., A.~P. McGovern, S.~J. Vollmer, and B.~A. Mateen (2021).
\newblock Improving survival of critical care patients with coronavirus disease
  2019 in england: a national cohort study, {M}arch to {J}une 2020.
\newblock {\em Critical care medicine\/}~{\em 49\/}(2), 209.

\bibitem[\protect\citeauthoryear{Flaxman, Mishra, Gandy, Unwin, Mellan,
  Coupland, Whittaker, Zhu, Berah, Eaton, et~al.}{Flaxman
  et~al.}{2020}]{flaxman2020lockdown}
Flaxman, S., S.~Mishra, A.~Gandy, H.~J.~T. Unwin, T.~A. Mellan, H.~Coupland,
  C.~Whittaker, H.~Zhu, T.~Berah, J.~W. Eaton, et~al. (2020).
\newblock Estimating the effects of non-pharmaceutical interventions on
  {COVID-19} in {E}urope.
\newblock {\em Nature\/}~{\em 584\/}(7820), 257--261.

\bibitem[\protect\citeauthoryear{Huang, Wang, Li, Ren, Zhao, Hu, Zhang, Fan,
  Xu, Gu, et~al.}{Huang et~al.}{2020}]{huang2020clinical}
Huang, C., Y.~Wang, X.~Li, L.~Ren, J.~Zhao, Y.~Hu, L.~Zhang, G.~Fan, J.~Xu,
  X.~Gu, et~al. (2020).
\newblock Clinical features of patients infected with 2019 novel coronavirus in
  {W}uhan, {C}hina.
\newblock {\em The {L}ancet\/}~{\em 395\/}(10223), 497--506.

\bibitem[\protect\citeauthoryear{Knock, Whittles, Lees, Perez~Guzman, Verity,
  Fitzjohn, Gaythorpe, Imai, Hinsley, Okell, Rosello, Kantas, Walters, Bhatia,
  Watson, Whittaker, Cattarino, Boonyasiri, Djaafara, Fraser, Fu, Wang, Xi,
  Donnelly, Jauneijaite, Laydon, White, Ghani, Ferguson, Cori, and
  Baguelin}{Knock et~al.}{2020}]{rep41}
Knock, E.~S., L.~K. Whittles, J.~A. Lees, P.~N. Perez~Guzman, R.~Verity, R.~G.
  Fitzjohn, K.~A.~M. Gaythorpe, N.~Imai, W.~Hinsley, L.~C. Okell, A.~Rosello,
  N.~Kantas, C.~E. Walters, S.~Bhatia, O.~J. Watson, C.~Whittaker,
  L.~Cattarino, A.~Boonyasiri, B.~A. Djaafara, K.~Fraser, H.~Fu, H.~Wang,
  X.~Xi, C.~A. Donnelly, E.~Jauneijaite, D.~J. Laydon, P.~J. White, A.~C.
  Ghani, N.~M. Ferguson, A.~Cori, and M.~Baguelin (2020).
\newblock Report 41: The {2020 SARS-CoV-2} epidemic in {E}ngland: key
  epidemiological drivers and impact of interventions.
\newblock {\em Imperial College London\/}.

\bibitem[\protect\citeauthoryear{Linton, Kobayashi, Yang, Hayashi,
  Akhmetzhanov, Jung, Yuan, Kinoshita, and Nishiura}{Linton
  et~al.}{2020}]{linton2020incubation}
Linton, N.~M., T.~Kobayashi, Y.~Yang, K.~Hayashi, A.~R. Akhmetzhanov, S.-m.
  Jung, B.~Yuan, R.~Kinoshita, and H.~Nishiura (2020).
\newblock Incubation period and other epidemiological characteristics of 2019
  novel coronavirus infections with right truncation: a statistical analysis of
  publicly available case data.
\newblock {\em Journal of clinical medicine\/}~{\em 9\/}(2), 538.

\bibitem[\protect\citeauthoryear{McAloon, Collins, Hunt, Barber, Byrne, Butler,
  Casey, Griffin, Lane, McEvoy, et~al.}{McAloon et~al.}{2020}]{McAloon20}
McAloon, C., {\'A}.~Collins, K.~Hunt, A.~Barber, A.~W. Byrne, F.~Butler,
  M.~Casey, J.~Griffin, E.~Lane, D.~McEvoy, et~al. (2020).
\newblock Incubation period of {COVID}-19: a rapid systematic review and
  meta-analysis of observational research.
\newblock {\em BMJ open\/}~{\em 10\/}(8), e039652.

\bibitem[\protect\citeauthoryear{Tay, Poh, R{\'e}nia, MacAry, and Ng}{Tay
  et~al.}{2020}]{tay2020trinity}
Tay, M.~Z., C.~M. Poh, L.~R{\'e}nia, P.~A. MacAry, and L.~F. Ng (2020).
\newblock The trinity of {COVID}-19: immunity, inflammation and intervention.
\newblock {\em Nature Reviews Immunology\/}, 1--12.

\bibitem[\protect\citeauthoryear{Verity, Okell, Dorigatti, Winskill, Whittaker,
  Imai, Cuomo-Dannenburg, Thompson, Walker, Fu, et~al.}{Verity
  et~al.}{2020}]{verity2020ifr}
Verity, R., L.~C. Okell, I.~Dorigatti, P.~Winskill, C.~Whittaker, N.~Imai,
  G.~Cuomo-Dannenburg, H.~Thompson, P.~G. Walker, H.~Fu, et~al. (2020).
\newblock Estimates of the severity of coronavirus disease 2019: a model-based
  analysis.
\newblock {\em The Lancet Infectious Diseases\/}.

\bibitem[\protect\citeauthoryear{Wang, Hu, Hu, Zhu, Liu, Zhang, Wang, Xiang,
  Cheng, Xiong, et~al.}{Wang et~al.}{2020}]{wang2020clinical}
Wang, D., B.~Hu, C.~Hu, F.~Zhu, X.~Liu, J.~Zhang, B.~Wang, H.~Xiang, Z.~Cheng,
  Y.~Xiong, et~al. (2020).
\newblock Clinical characteristics of 138 hospitalized patients with 2019 novel
  coronavirus--infected pneumonia in {W}uhan, {C}hina.
\newblock {\em Jama\/}~{\em 323\/}(11), 1061--1069.

\bibitem[\protect\citeauthoryear{Wood}{Wood}{2017}]{wood2017igam}
Wood, S.~N. (2017).
\newblock {\em Generalized Additive Models: An Introduction with R\/} (2 ed.).
\newblock Boca Raton, FL: CRC press.

\bibitem[\protect\citeauthoryear{Wood}{Wood}{2020}]{wood2020arxiv}
Wood, S.~N. (2020).
\newblock Did {COVID}-19 infections decline before {UK} lockdown?
\newblock {\em arXiv preprint ArXiv:2005.02090\/}.

\bibitem[\protect\citeauthoryear{Wu, Leung, Bushman, Kishore, Niehus,
  de~Salazar, Cowling, Lipsitch, and Leung}{Wu et~al.}{2020}]{wu2020covid}
Wu, J.~T., K.~Leung, M.~Bushman, N.~Kishore, R.~Niehus, P.~M. de~Salazar, B.~J.
  Cowling, M.~Lipsitch, and G.~M. Leung (2020).
\newblock Estimating clinical severity of {COVID}-19 from the transmission
  dynamics in {W}uhan, {C}hina.
\newblock {\em Nature Medicine\/}~{\em 26\/}(4), 506--510.

\bibitem[\protect\citeauthoryear{Zhou, Yu, Du, Fan, Liu, Liu, Xiang, Wang,
  Song, Gu, et~al.}{Zhou et~al.}{2020}]{zhou2020clinical}
Zhou, F., T.~Yu, R.~Du, G.~Fan, Y.~Liu, Z.~Liu, J.~Xiang, Y.~Wang, B.~Song,
  X.~Gu, et~al. (2020).
\newblock Clinical course and risk factors for mortality of adult inpatients
  with {COVID}-19 in {W}uhan, {C}hina: a retrospective cohort study.
\newblock {\em The {L}ancet\/}~{\em 395\/}(10229), 1054--1062.

\end{thebibliography}
\end{document}